\documentclass[11pt]{article}

\usepackage{jheppub}
\usepackage{amssymb,amsmath,latexsym,bm,amsfonts}

\usepackage{longtable}
\usepackage{color,xcolor}
\usepackage{indentfirst}
\usepackage{mathtools,tensor,graphicx}

\usepackage{enumitem}
\usepackage{pifont}

\usepackage{cleveref}
\usepackage{dsfont}
\usepackage{enumitem}
\usepackage{braket}
\usepackage{slashed}
\usepackage{float}
\usepackage{caption}
\usepackage{subfigure}
\usepackage{makecell}
\usepackage{epsfig}

\newcommand{\bea}{\begin{eqnarray}}
\newcommand{\eea}{\end{eqnarray}}
\newcommand{\be}{\begin{equation}}
\newcommand{\ee}{\end{equation}}
\newcommand{\ba}{\begin{align}}
\newcommand{\ea}{\end{align}}
\newcommand{\ben}{\begin{enumerate}}
\newcommand{\een}{\end{enumerate}}
\newcommand{\bi}{\begin{itemize}}
\newcommand{\ei}{\end{itemize}}

\newcommand{\p}{\partial}

\title{Higher-derivative supersymmetric effective field theories}

\author[a,c]{Osmin Lacombe,}
\author[b]{Lorenzo Paoloni,}
\author[a,c]{Francisco G. Pedro}
\affiliation[a]{Dipartimento di Fisica e Astronomia, Universit\'a di Bologna, via Irnerio 46, 40126 Bologna, Italy}
\affiliation[b]{Instituto de F\'isica Te\'orica IFT-UAM/CSIC,
C/ Nicol\'as Cabrera 13-15, Campus de Cantoblanco, 28049 Madrid, Spain}
\affiliation[c]{INFN, Sezione di Bologna, viale Berti Pichat 6/2, 40127 Bologna, Italy}

\emailAdd{deriusosmin.lacombe@unibo.it}
\emailAdd{lorenzo.paoloni@csic.es}
\emailAdd{francisco.soares@unibo.it}

\abstract{In this paper we study higher-derivative supersymmetric effective field theories focusing on the systematic procedure for the elimination of ghosts from the spectrum. Particular attention is paid to the auxiliary fields, for which the higher-derivative terms induce non-algebraic equations of motion. By employing field redefinitions or the reduction of order procedure (both in component and superfield language) we show that the auxiliary fields remain non-dynamical in the EFT and that on shell they give rise to both derivative and non-derivative corrections to the scalar action. These methods are applied to the search for a SUSY embedding of the  DBI action and to the dimensional reduction of HD terms for the K\"ahler moduli in type IIB string compactifications.
}

\begin{document} 
\maketitle
\flushbottom

\newpage

\section{Introduction}\label{intro}

 Actions with higher-derivative (HD) operators generically lead to higher-order equations of motion, which require the specification of additional initial conditions for the solution of the Cauchy problem and hence signal the presence of new degrees of freedom. Such new degrees of freedom can be described by new fields called ghosts, featuring the wrong sign kinetic term, and imply the presence of the Ostrogradsky's instability \cite{Ostrogradsky:1850fid}. There are however two instances in which HD theories are ghost-free. Firstly, it is well know that not all HD actions lead to equations of motion that are of higher order in time derivatives. Scalar-tensor theories with HD operators and two-derivative equations of motion (hence free of ghosts) have been classified in \cite{Horndeski:1974wa}. The simplest examples are $P(X\equiv (\partial \phi)^2)$ theories. Other theories with HD operators lead to higher-order equations of motion but avoid the presence of ghosts due to particular degeneracy conditions \cite{Langlois:2015cwa}. The second situation where a theory can be free of instabilities associated with the presence of HD terms is when it is taken to be an effective field theory (EFT), valid below a cutoff energy scale. In this case the HD terms can be recast as perturbative corrections to a well defined two-derivative action.

Higher-derivative operators naturally arise from certain UV complete theories such as string theory, the DBI action being the prime example. A large corner of 4d reductions of string theory preserves some amount of supersymmetry (SUSY), so that the study of HD operators in supersymmetric theories is rather natural. One can  argue that when the UV theory is consistent, the apparent instabilities in the EFT are a consequence of the truncation at finite order of an entire series of HD operators. Such series are  free from instabilities but their truncations are not, see e.g. \cite{Buchbinder:1994iw,Pickering:1996he} for concrete examples. 
String theory compactifications to four dimensions are described at leading order by two-derivative SUSY action and feature HD operators suppressed by powers of the cutoff scale, thus only contributing as perturbations. Following this discussion, it is  necessary to understand how to deal with HD perturbations to well-defined two-derivative supersymmetric EFT. We will see how a consistent treatment of the HD terms relegates their associated instabilities to energies above the cutoff of the EFT, leaving a healthy and well defined theory at low energies \cite{Burgess:2014lwa}. Our work builds on and extends \cite{Khoury:2010gb,Koehn:2012ar,Koehn:2012np,Farakos:2012qu,Ciupke:2015msa} where supersymmetric extensions of ghost-free scalar field theories and some of their phenomenological consequence have been studied.

The goal of the present paper is to shed light on the treatment of higher-derivative perturbations to supersymmetric EFTs. Our main message is that one can consider any HD SUSY operator and render the EFT consistent below its cutoff by employing the same EFT methods used in the absence of SUSY. More concretely, we focus on chiral SUSY theories and review how most HD operators induce derivative terms for the auxiliary field. These generally lead to non-algebraic equations of motion which naively seem to render these fields dynamical. Whereas previous works have avoided such operators or simply set derivatives of the auxiliary fields to zero, we show how to systematically eliminate them together with the scalar ghost degrees of freedom by employing methods equivalent to those used in e.g. \cite{Antoniadis:2007xc,Antoniadis:2008es,Antoniadis:2009rn} for particular operators. To do so, we can either generalise the method of \cite{Jaen:1986iz} to the supersymmetric theories or use field redefinitions. These methods eliminate the derivatives of the auxiliary field without explicitly solving its equations of motion (contrary to e.g. \cite{Antoniadis:2019xwa} where the auxiliary field's equations of motion, containing derivatives, were solved perturbatively before performing the field redefinition). After proper treatment of HD operators, one obtains a two-derivative, ghost-free effective action where kinetic and interaction terms receive perturbative corrections that can be traced back to the HD operators and that can lead to interesting phenomenological consequences.

The paper is organised as follows. In \Cref{sec:HD_fieldtheory}, we study higher-derivative operators in quantum field theories for scalar fields. We start by reviewing how these operators generically lead to instabilities signalled by the presence of ghost fields. We then show in detail how to systematically deal with such HD operators to eliminate potential instabilities from the EFT. In \Cref{sec:HDSUSY} we extend this analysis to the case of supersymmetric EFTs. In that case, particular attention is paid to the auxiliary fields, which naively seem to become dynamical in the presence of HD operators for scalar superfields. We analyse in detail, both in components and in superspace, four-derivative operators correcting the scalar superfield K\"ahler potential. We apply these EFT methods in \Cref{sec:Applications} to three phenomenologically interesting cases. We first review higher-derivative extensions of the MSSM, we then address the scalar sector of the supersymmetric Dirac-Born-Infeld action describing the brane position moduli. We conclude with a preliminary study of higher-derivative operators of 4d $\mathcal{N}=1$ SUSY reductions of type  IIB string theory and their implication for moduli stabilisation. Supplementary material is presented in the three appendices. \Cref{AppendixQuantization} contains a short review of the quantisation of theories with higher-derivative operators, while \Cref{AppendixSuperfields} shows our SUSY conventions and notations. \Cref{HDMSSMcomponents} shows in detail the elimination of dimension-five HD operator in components, showcasing explicitly the discrepancy arising when incorrectly applying the procedure for the elimination of HD operators from the  EFT.
 
\section{Higher-derivative operators in scalar field theories} \label{sec:HD_fieldtheory}

In this section we review the problems associated with HD actions and then employ two equivalent techniques, reduction of order in the equation of motion (eom) in \ref{JLMconstraints} and field redefinitions in the action in \ref{fieldredefs}, to eliminate HD terms in scalar (quantum) field theories.   
The methods introduced here will then be used in section \ref{sec:HDSUSY} to consistently treat HD terms in supersymmetric field theories.
 
\subsection{Fundamental higher-derivative operators}\label{basicExampleHDtheory}
 
Before dealing with the treatment of HD terms in EFTs, let us start by analysing a simple example of genuine HD theory (by which we mean a theory that is truly HD instead of an EFT with HD terms) to illustrate the problems that generically arise in such cases. We consider the theory of a real scalar field $\phi$, described by the following Lagrangian\footnote{This example is the field theory equivalent of the point-particle example treated in \cite{Solomon:2017nlh}.}:
\begin{equation}
\mathcal{L}=-\frac{1}{2}\partial^{\mu}\phi\partial_\mu \phi-\frac{1}{2}m^{2}\phi^{2}-\frac{1}{2M^2}(\Box \phi)^{2}. \label{e1}
\end{equation}
The mass scale $M$ guarantees the correct dimension for the higher-derivative term $(\Box\phi)^{2}$ and will later be related to the expansion coefficient $\epsilon$ in the EFT context, through $\epsilon\equiv\frac{1}M.$

The quantisation procedure of $\phi$ is shown in \cref{AppendixQuantization}, here  we restrict our attention to the equation of motion derived from \eqref{e1}.  The Euler-Lagrange equation for higher-derivative theories, derived by requiring stationarity of the action, is of the form\footnote{We adopt the notation $(\partial_\mu)^k\equiv\partial_{\mu_1 \mu_2...\mu_k}$.}:
\begin{equation}
\sum_{k=0}^n (-1)^k (\partial_\mu)^k \frac{\partial \mathcal{L}}{\partial ((\partial_\mu)^k \phi) }=0 . \label{eomHD}
\end{equation}
 For the Lagrangian \eqref{e1} it simply reads:
\begin{equation}
\left(-\Box+m^{2}+\epsilon^2 \Box^{2}\right)\phi=0, \label{e5}
\end{equation}
and leads to the fourth order energy-momentum relation:
\begin{equation}
\epsilon^{2}p^{4}+p^{2}+m^{2}=0,
\end{equation}
which admits two distinct solutions:
\begin{align}
-p_{c}^{2}&\equiv\omega_{c\vec{p}}^{2}-|\vec{p}|^{2}=\frac{1-\sqrt{1-4m^{2}\epsilon^{2}}}{2\epsilon^{2}}=m^{2}+m^{4}\epsilon^{2}+O(\epsilon^{4}), \label{momentumc}\\
-p_{d}^{2}&\equiv\omega_{d\vec{p}}^{2}-|\vec{p}|^{2}=\frac{1+\sqrt{1-4m^{2}\epsilon^{2}}}{2\epsilon^{2}}=\frac{1}{\epsilon^2}-m^{2}-m^{4}\epsilon^{2}+O(\epsilon^{4}). \label{momentumd}
\end{align}
These two modes have convergent or divergent momentum respectively in the limit $\epsilon\rightarrow0$ (or $M\rightarrow\infty$). In this limit, the Lagrangian \eqref{e1} recovers the standard Lagrangian of a massive scalar field.  The divergence of $p_d^2$ shows that the mass of the associated mode goes to infinity. 

The quantisation of $\phi$, shown in \cref{AppendixQuantization}, indeed shows that the theory contains two types of particles. The ``convergent-type" and  ``divergent-type" particles of momentum $\vec{p}$, which contribute $+ \sqrt{1-4m^{2}\epsilon^{2}}\;\omega_{c\vec{p}}$ and $- \sqrt{1-4m^{2}\epsilon^{2}}\;\omega_{d\vec{p}}$ to the energy of system. Higher-derivative terms thus introduce negative energy states, or equivalently negative norm states if one modifies the commutation relations. As explained in  \cref{AppendixQuantization}, divergent-type particles also lead to an energy that is unbounded from below, thus to a system without a ground state.

Similar conclusions regarding the theory of \eqref{e1} can be found by introducing a new field variable and diagonalising the problem \cite{deUrries:1998obu}. In this simple example, one can simultaneously diagonalise both the kinetic and mass matrices of the two-scalar field system (this is in general not possible). Let us define the dimensionless constants:
\begin{align}
\mu_{c}&\equiv\frac{1-\sqrt{1-4m^{2}\epsilon^{2}}}{2}= -p_{c}^{2}\epsilon^{2}=m^{2}\epsilon^{2}+O(\epsilon^{4}), \label{muc}\\
\mu_{d}&\equiv\frac{1+\sqrt{1-4m^{2}\epsilon^{2}}}{2}= -p_{d}^{2}\epsilon^{2}=1-m^{2}\epsilon^{2}+O(\epsilon^{4}), \label{mud}
\end{align}
which obey the following identities:
\begin{align}
&\mu_{c}+\mu_{d}=1, \qquad \mu_{c}\mu_{d}=m^{2}\epsilon^{2}=\frac{m^2}{M^2}, \qquad \mu_{c}^{2}=\mu_{c}-m^{2}\epsilon^{2}, \qquad\mu_{d}^{2}=\mu_{d}-m^{2}\epsilon^{2}.
\end{align}
Defining the two scalar fields:
\begin{align}
\varphi_{1}&=-\frac{\mu_{c}}{\sqrt{\mu_{d}-\mu_{c}}}\phi+\frac{1}{M^{2}\sqrt{\mu_{d}-\mu_{c}}}\Box\phi,\label{e35}\\
\varphi_{2}&=\frac{\mu_{d}}{\sqrt{\mu_{d}-\mu_{c}}}\phi-\frac{1}{M^{2}\sqrt{\mu_{d}-\mu_{c}}}\Box\phi,\label{e36}
\end{align}
brings the Lagrangian of \cref{e1}  to the form:
\begin{equation}
\mathcal{L}=-\frac{1}{2}\varphi_{1}(\Box-\mu_{d}M^{2})\varphi_{1}+\frac{1}{2}\varphi_{2}(\Box-\mu_{c}M^{2})\varphi_{2}. \label{e37}
\end{equation}
As anticipated, the Lagrangian has diagonal kinetic and mass terms for the two scalars. This allows to directly identify the physical modes and it is clear that the kinetic term of the field $\varphi_{1}$ has the wrong sign: the latter is a ghost. Extracting its mass-momentum relation $-p_{1}^{2}=\mu_{d}M^{2}$ one deduces that the ghost $\varphi_{1}$ is the divergent-type particle.
Similar considerations show that $\varphi_{2}$ is the convergent-type particle. 

\subsection{Treatment of higher-derivative operators in EFTs}

We now review how instabilities of HD theories can be treated in effective field theories studied below a certain cutoff scale $M$. In this context, higher-derivative interactions should be thought as low-energy remnants of high-energy interactions. The latter involve UV degrees of freedom, above the scale $M$, that have been integrated out to obtain the effective theory with HD operators. Ghost fields coming from these HD operators might thus be artefacts of the effective theory below the scale $M$, see \cref{intro}.

We show below how to deal with HD operators in effective theories with unknown UV completion. In this framework, HD operators are treated as perturbative corrections to a well-defined two-derivative theory, suppressed by powers of $1/M$. They are thus included at least at order $O(\epsilon)$ in the expansion of the Lagrangian in powers of $\epsilon\equiv1/M$. 

Consider the HD theory of a scalar field and expand its Lagrangian in $\epsilon$ as
\begin{equation}
\mathcal{L}=-\frac{1}{2}\partial^{\mu}\phi\partial_{\mu}\phi -\sum_{k=0}^{n}\epsilon^{k}V_{k}(\phi,\partial\phi,\dots,\partial^{(k)}\phi)+O(\epsilon^{n+1}). \label{e40}
\end{equation}
 By writing the Lagrangian in this form we have separated the canonical kinetic term from the scalar potential and higher-derivative interactions. In this notation $V_0(\phi)$ is the scalar potential, $V_1(\phi,\partial\phi)$ contains corrections to the scalar potential and the canonical kinetic term, while $V_k(\phi,\partial\phi,\dots,\partial^{(k)}\phi)$ for $k\geq 2$ encode higher-derivative interactions. The above Lagrangian is seen as an expansion in the order parameter $\epsilon$ and considered as an approximation to order $O(\epsilon^{n+1})$. 
 
 In what follows, we analyse the resulting EFT using two equivalent methods to exorcise the ghosts (reduction of order/JLM and field redefinitions), showing that there are no instabilities when the HD terms are treated as a perturbative correction to the well-defined two-derivative theory.
 
\subsubsection{Reduction of order/ JLM procedure} \label{JLMconstraints}

The method developed by Jaen, Llosa and Molina (henceforth called JLM/reduction of order) in \cite{Jaen:1986iz} is a systematic procedure for reducing the differential order of the equations of motion obtained from a HD Lagrangian interpreted as an EFT.

The equations of motion derived from the  HD Lagrangian \eqref{e40} will be considered as a perturbative expansion up to order $O(\epsilon^{n+1})$:
\begin{equation}
\Box \phi-\sum_{k=0}^{n}\epsilon^{k}F_{k}(\phi,\dots,\partial^{(2k)}\phi)=O(\epsilon^{n+1}), \label{e41}
\end{equation} 
where
\begin{equation}
F_{k}\equiv\sum_{r=0}^{k}(-1)^{r}{\partial_{\alpha_1}\partial_{\alpha_2}\dots\partial_{\alpha_r}}\frac{\partial V_{k}}{\partial (\partial_{\alpha_1}\partial_{\alpha_2}\dots\partial_{\alpha_r}\phi)}, \label{Ffunctions}
\end{equation}
encodes the HD terms for $k\ge 2$.

The goal of the JLM procedure is to obtain an equation of motion without higher-derivative terms. The solution of this equation of motion describes the same physics as the ones of \eqref{e41}, which includes  HD operators. It is thus a systematic way to treat and eliminate HD terms at the level of the equations of motion of the system. 

The procedure relies on constraints to eliminate from the EFT the HD terms that would otherwise give rise to unphysical degrees of freedom. These constraints are derived from \eqref{e41} by working order by order in the expansion parameter $\epsilon$ and allow to express each HD term in terms of lower derivative  ones. The procedure is iterative and starts with a  derivation of the constraints needed to eliminate the $\epsilon^n$ HD terms included in $F_n$, and goes down order by order until $\mathcal{O}(\epsilon)$. In the end, only terms without higher derivatives remain in the final equation of motion. 

\paragraph{The algorithmic procedure} Let us now explain the two different steps of the JLM procedure, performed for each order of approximation $O(\epsilon^k)$ with $n\geq k \geq 1$. At order $k$, they are:
\begin{itemize}
\item[$\triangleright$]First, derive the set of constraints needed to eliminate $O(\epsilon^k)$ higher-derivative terms included in $F_k$. These constraints are obtained by multiplying the full equation of motion \eqref{e41} and its space-time derivatives by  $\epsilon^{k}$,  and then truncating at order $O(\epsilon^{n})$. They thus read:
\begin{equation}
\forall  s \leq 2k-2, \quad  \epsilon^k \partial^{(s)} \Box \phi=\sum_{j=0}^{n-k} \epsilon^{k+j} \partial^{(s)} F_{j} + O(\epsilon^{n+1})\ . \label{constraints}
\end{equation}
For the moment, we omit Lorentz indices for the derivatives, generically denoted $\partial^{(s)}$.
We  come back to the accurate treatment of contractions below. 
\item[$\triangleright$] Second, take care of the higher-derivative terms on the right-hand sides of the above constraints. These terms are eliminated using the previous constraints, obtained either at the same order with less derivatives, for $j=0$, or at the previous order for $j> 0$.  For instance, the $\epsilon^{k+j}\partial^{(s)}F_{j}(\phi)$ terms of the constraints \eqref{constraints}, with $s\geq 2$, depend {\it a priori} on $\epsilon^{k}\partial^{(s)}\phi$. They can be treated through the constraints of higher orders.
\end{itemize}
Iterating this algorithm progressively lowers the differential order of the equation of motion and eventually leads to the final equation of motion of a two-derivative theory:
\begin{equation}\Box\phi=\sum_{k=0}^{n}\epsilon^{k}\bar{F}_{k}(\phi,\partial\phi,\partial^{(2)}\phi)+O(\epsilon^{n+1})\,.\label{e43}
\end{equation}
The $\bar{F}_{k}$ are the result of the repeated substitutions described for each step, properly grouped in terms of powers of $\epsilon$. Each $\bar{F}_{k}$ will generically depend on all the initial $F_i$ for $i\leq k$.
Note that \cref{e43} is compatible with all the previous constraints, properly multiplied by powers of $\epsilon$, differentiated, and plugged back to eliminate possible undesired higher derivatives.

We should note that the reduction of order/JLM procedure applied to field theories is more involved that its application to mechanics as: 1) there are two non-equivalent operators containing two time derivatives, namely $\Box \phi$ and $\partial_{\mu \nu}\phi$, and 2)
not all HD operators can be written as spacetime derivatives of the leading order eom. Point 1) is the reason for the $\partial^{(2)}\phi$ dependence in the rhs of \cref{e43}, though we stress that the eom are still of order two in time derivatives.
Formally only higher-order time derivatives give rise to ghosts and need to be eliminated. In field theory, this will sometimes require performing a 1+3 splitting of the Lorentz invariant theory to eliminate higher-order time derivatives. More concretely, in the scalar field case, not all HD operators are given as derivatives of $\Box\phi$, e.g. while  a term of the form $\partial_\mu\Box\phi$ can be immediately eliminated form the eom,  a term proportional to $\partial_{\mu}\partial_{\nu}\partial_{\rho}\phi$ cannot be replaced without first isolating the third order time derivative terms \cite{Solomon:2017nlh}.

Let us now comment on the use of the JLM procedure, introduced in \cite{Jaen:1986iz} in mechanics and generalised above to field theory. This algorithmic procedure is based on the manipulation of the equation of motion of a HD theory, written in terms of an expansion parameter $\epsilon$. It produces a two-derivative equation of motion with the same solutions as the initial one, at a certain total order $O(\epsilon^n$). In the EFT context, this expansion is natural below a certain cutoff scale $M$ and the procedure thus leads to a two-derivative effective equation of motion, describing the same physics up to a certain order. One can try and reconstruct an effective Lagrangian leading to this two-derivative equation of motion. This is not necessary to find the solutions to the dynamical problem, but can be very useful for physical interpretation and to compare with the original HD Lagrangian. As explained below, in some cases this effective Lagrangian can be obtained using directly the effective equation of motion in the original Lagrangian to eliminate HD terms, a procedure that must be done with extreme care as it is not correct in general.

\paragraph{Application of the procedure for the first two orders}  Let us now show more explicitly how to apply the steps for the first two orders, $k=n, n-1$. We thus want to eliminate the  HD terms $\epsilon^n F_n$ and  $\epsilon^{n-1} F_{n-1}$ of the equation of motion \eqref{e41}. 

The first step consist in deriving the constraints, at order $k=n$, necessary to treat $F_n$. These constraints are obtained by multiplying \cref{e41} by $\epsilon^n$ and will thus  only contain the $\epsilon^0$ terms of \cref{e41}, $\Box \phi$ and $F_0$, and their derivatives. Indeed, according to \cref{constraints} they are:
\begin{align}
\epsilon^{n}&\Box \phi=\epsilon^{n}F_{0}(\phi)+O(\epsilon^{n+1}),\nonumber \\
&\vdots  \label{JLMfirststep} \\
\epsilon^{n}&\partial^{(2n-2)}\Box\phi=\epsilon^{n}\partial^{(2n-2)}F_{0}(\phi)+O(\epsilon^{n+1})\,\,. \nonumber
\end{align}
The second step of the procedure consists in eliminating the higher-order terms on the rhs of the constraints. For instance, the constraint obtained at order $k=n$ by taking two derivatives, given implicitly in \eqref{JLMfirststep}, reads
\begin{align}
\epsilon^{n}\partial^{(2)}\Box\phi&=\epsilon^{n}\partial^{(2)}F_{0}(\phi)+O(\epsilon^{n+1})\equiv\epsilon^{n}f(\phi,\partial\phi,\Box\phi)+O(\epsilon^{n+1})\nonumber\\
&=\epsilon^{n}f(\phi,\partial\phi,F_{0})+O(\epsilon^{n+1})\equiv\epsilon^{n}g(\phi,\partial\phi)+O(\epsilon^{n+1}), \label{constraint4}
\end{align}
with $f$ involving the derivatives of $F_0$. In the second line we have used the expression for $\Box \phi$ obtained from the first line of \cref{JLMfirststep}.  This process can be  iterated for all the constraints obtained at order $k=n$ with higher-derivative terms $\epsilon^{n}\partial^{(s)}F_{0}(\phi)$ with $s\geq 2$. 

Once $F_{n}$ has been exorcised from HD terms through substitutions of the above constraints, one can go to the next order and treat $\epsilon^{n-1}F_{n-1}$, which depends on $\phi, \partial\phi, \ldots, \partial^{(2n-2)}\phi$. According to \cref{constraints}, the necessary constraints at order $k=n-1$ are thus of the form:
\begin{align}
\epsilon^{n-1}&\Box\phi=\epsilon^{n-1}F_{0}(\phi) + \epsilon^{n}F_{1}(\phi,\partial\phi,\partial^{(2)}\phi)+O(\epsilon^{n+1}),\nonumber\\
&\vdots \\
\epsilon^{n-1}&\partial^{(2n-4)}\Box\phi= \epsilon^{n-1}\partial^{(2n-4)}F_{0}(\phi) + \epsilon^{n}\partial^{(2n-4)}F_{1}(\phi,\partial\phi,\partial^{(2)}\phi)+O(\epsilon^{n+1}). \nonumber
\end{align}
 Substitutions similar to the ones for the $\epsilon^n$ constraints are again required to keep only $\phi$, $\partial\phi$, and $\partial^{(2)}\phi$ dependences throughout the various differentiations. After being treated this way, these $\epsilon^{n-1}$ constraints can eventually be used in $\epsilon^{n-1}F_{n-1}$.

\subsubsection{Field redefinitions in the Lagrangian}\label{fieldredefs} 

A well-known method to obtain on-shell equivalent theories consist in applying field redefinitions at the level of the Lagrangian \cite{Georgi:1991ch,Weinberg:2008hq, Burgess:2007pt, Grosse-Knetter:1993tae,Gong:2014rna}. When applied in the context of effective theories, it allows for the elimination of some higher-order derivative terms, when they are proportional to the lowest order equations of motion. We briefly review how field redefinitions operate in our case. For clarity, let us rewrite here the Lagrangian of \cref{e40}:
\begin{align}
\mathcal{L}=&-\frac{1}{2}\partial^{\mu}\phi\partial_{\mu}\phi -\sum_{k=0}^{n}\epsilon^{k}V_{k}(\phi,\partial\phi,\dots,\partial^{(k)}\phi)+O(\epsilon^{n+1}).
\end{align}
The use of field redefinitions allows to eliminate terms proportional to the lowest order equation of motion, order by order. For a certain order $O(\epsilon^{k})$ one should thus write them in the form
\begin{equation}
\mathcal{L}=\mathcal{L}_{k-1}-\epsilon^{k}T_{k}\Box\phi+\epsilon^{k}\tilde{V}_{k}+O(\epsilon^{k+1}), \label{LagrangianexpansionBox}
\end{equation}
by defining
\begin{equation}
 \tilde{V}_{k}\equiv V_{k}+T_{k}\Box\phi.
 \end{equation} 
 We introduced  $T_{k}$, which is thus a function of $\phi$ and its derivatives. We can always extract a term proportional to $\Box\phi$ from the $O(\epsilon^k)$ term by integrating by parts repeatedly a generic term $f(\phi,\partial\phi,\dots)\partial^{(j)}\phi$, except for constant $f$ where such term is a total derivative.

  It is  possible to perform the field redefinition
\begin{equation}
\phi\longrightarrow \phi+\delta\phi=\phi + \epsilon^{k} \, T_{k},  \label{fieldredef1}
\end{equation}
under which the Lagrangian transforms as
\begin{align}
\mathcal{L}\longrightarrow \quad  \mathcal{L}  + \delta \phi \frac{\delta \mathcal L}{\delta \phi} + \frac{1}{2} (\delta \phi )^2 \frac{\delta^2 \mathcal L}{\delta \phi^2 } + \dots  + \frac{1}{p!} (\delta \phi )^p \frac{\delta^p \mathcal L}{\delta \phi^p }  +  O(\epsilon^{n+1}). \label{redefinedLagrangian}
\end{align}
The first-order perturbation of the Lagrangian under the field redefinition is exactly the field equation \eqref{e41} and reads:
\begin{equation}
 \frac{\delta \mathcal L}{\delta \phi}= \Box\phi - \sum_{k=0}^{n}\epsilon^{k}F_{k}(\phi,\dots,\partial^{(2k)}\phi) + O(\epsilon^{n+1}). \label{firstorderpert}
\end{equation}
This tells us that the field redefinition \eqref{fieldredef1} will always allow to eliminate HD terms proportional to $\Box\phi$ at order $\epsilon^k$. Due to the higher order terms in \cref{fieldredef1,redefinedLagrangian}, such field redefinition will also generate HD terms at higher orders, that can be treated iteratively. Hence the field redefinition method starts from the lowest order in  $\epsilon$ and proceeds by increasing the order, until the elimination of all HD terms at order $O(\epsilon^{n+1})$ has been accomplished.

The fact that we work in an $\epsilon$ expansion appears in two ways. The field redefinition \eqref{fieldredef1} includes an $\epsilon^k$ dependence, while the resulting Lagrangian is considered at order $\epsilon^{n+1}$. Hence, one only has to perturb the Lagrangian \eqref{redefinedLagrangian} at order $p$ such that $k p < n+1$. 

This last consideration tells us that, for a field redefinition proportional to $\epsilon^n$ with $n\geq 1$, it is sufficient to look at the first order perturbation of the Lagrangian. Under \cref{fieldredef1} with $k=n$, the Lagrangian \eqref{LagrangianexpansionBox} hence goes to:
\begin{align}
\mathcal{L}\longrightarrow&   \quad  \mathcal{L}  + \delta \phi \frac{\delta \mathcal L}{\delta \phi} + O(\epsilon^{n+1}) =\mathcal{L} + \epsilon^{n}T_{n}\left(\frac{\partial\mathcal{L}_{0}}{\partial\phi}-\partial_{\alpha}\frac{\partial\mathcal{L}_{0}}{\partial(\partial_{\alpha}\phi)}\right)+O(\epsilon^{n+1})\nonumber\\
					  &=\mathcal{L} + \epsilon^{n}T_{n}\left(-\frac{\partial V_{0}}{\partial\phi} + \Box\phi\right)+O(\epsilon^{n+1})=\mathcal{L}_{n-1}+\epsilon^{n}\tilde{V}_{n}-\epsilon^{n}T_{n}\frac{\partial V_{0}}{\partial\phi}+O(\epsilon^{n+1})\,\,.
\end{align}
The field redefinition thus removes the  $\epsilon^{n}T_n\Box\phi$ term and replaces it using the zeroth order eom $\Box\phi=V_0'$. Other terms of the same form could remain in $T_{n}$. They can be eliminated by integrating by parts, if necessary, to create other $\epsilon^{n}\Box\phi$ terms, and making a new field redefinition of the same form. Iterating the procedure at the order $k=n$ leads to:
\begin{equation}
\mathcal{L}=\mathcal{L}_{n-1}+\epsilon^{n}\bar{V}_{n}(\phi,\partial\phi)\label{e49},
\end{equation}
where all higher-derivative terms have been removed.

\paragraph{Using JLM constraints in the Lagrangian} We see that this first field redefinition is completely equivalent to substituting the first JLM constraints, proportional to $\epsilon^{n}$, directly into the Lagrangian. This is due to the  fact that the JLM constraints are derived from the eom and that at the order we considered the field redefinition replaces the HD terms by powers of the first order eom. This equivalence actually holds true only when the first order perturbation in \cref{redefinedLagrangian} is sufficient, hence at all orders $O(\epsilon^k)$ such that $2 k \geq n+1$. When this is not the case one cannot directly replace the constraints obtained by the JLM procedure in the Lagrangian. The equivalence is thus trivial for the $n=1$ case, for which one only studies the effect of $O(\epsilon)$ corrections up to $O(\epsilon^2)$ effects, see \cite{Grosse-Knetter:1993tae}.

\subsection{A simple EFT example  with higher-derivative operators}\label{JLMfields}

In this section we revisit the simple example of \cref{basicExampleHDtheory}, but we now interpret it as an EFT, assuming that $m\ll M$ or yet $\epsilon m \ll1$. The Lagrangian \eqref{e1} at order $O(\epsilon^2)$ reads: 
\begin{equation}
\mathcal{L}=-\frac{1}{2}(\partial\phi)^{2}-\frac{1}{2}m^{2}\phi^{2}-\frac{1}{2}\epsilon^{2}(\Box\phi)^{2}+O(\epsilon^{3})\,\,.\label{e55}
\end{equation}
from which we find the full equation of motion:
\begin{equation}
\Box\phi-m^{2}\phi-\epsilon^{2}\Box^{2}\phi=O(\epsilon^{3})\label{e56}\ .
\end{equation}
Comparison with \eqref{e41} prompts the identifications $F_{0}=m^{2}\phi$, $F_{1}=0$ and  $F_{2}=\Box^2\phi$, for the functions introduced in \cref{Ffunctions}. Following the JLM procedure, the constraints at order $k=2$ are obtained by multiplying the above equation and its appropriate derivatives by $\epsilon^{2}$ and truncating. For the equation without derivatives, we obtain:
\begin{equation}
\epsilon^{2}\Box\phi=\epsilon^{2}m^{2}\phi+O(\epsilon^{3})\,\,.\label{e57}
\end{equation}
We also need a constraint for $\epsilon^{2}\Box^2\phi$ in \eqref{e56}, so that we should differentiate twice and contract \eqref{e57} to obtain the second constraint:
\begin{equation}
\epsilon^{2}\Box^2\phi=\epsilon^{2}m^{2}\Box \phi+O(\epsilon^{3}).
\end{equation}
The second step of the JLM procedure eliminates the $\Box\phi$ dependence in the rhs by making use of the previous constraint, namely \cref{e57}. It thus amounts to writing:
\begin{equation}
\epsilon^{2}\Box^2\phi = \epsilon^2 m^4 \phi+ O(\epsilon^3)\ . \label{e58}
\end{equation}
Hence, the healthy equation of motion is the one obtained by plugging \eqref{e58} into the original equation \eqref{e56}: 
\begin{equation}
\Box\phi=m^{2}(1+m^{2}\epsilon^{2})\phi+O(\epsilon^{3}).\label{e59}
\end{equation}
We can thus identify the $\bar F_k$ functions defined in \cref{e43} to $\bar{F}_{0}=m^{2}\phi$, $\bar{F}_{1}=0$ and  $\bar{F}_{2}=m^{4}\phi$.  This equation of motion is clearly of second order and thus well behaved.
As mentioned above, one can try and infer a Lagrangian leading to this equation of motion. It reads:
\begin{equation}
\mathcal{L}=-\frac{1}{2}\partial^{\mu}\phi\partial_{\mu}\phi-\frac{1}{2}m^{2}(1+m^{2}\epsilon^{2})\phi^{2}+O(\epsilon^{3})\label{e60}\ .
\end{equation}
We also notice that using \eqref{e57} twice directly in the original Lagrangian produces the same equation of motion. This is not a surprise since the constraints are used to treat a term proportional to $\epsilon^2$ at order $O(\epsilon^3)$ and that $2\times2 \geq 3$. As explained below \cref{e49}, in that case plugging the JLM constraints directly in the Lagrangian is equivalent to making field redefinitions and is thus fully justified.

The Lagrangian \eqref{e60} is just the one for a free real scalar field of mass $m^{2}(1+m^{2}\epsilon^{2})$. Treating the HD terms as perturbative corrections, i.e. considering  the regime $\epsilon^2 m^2 \ll 1$, thus amounts to correcting the mass of the real scalar field $\phi$. It is easy to make the connection with the way we dealt with this case in \cref{basicExampleHDtheory}, by comparing to the Lagrangian \eqref{e37} obtained after the explicit introduction of a new degree of freedom. Inserting the constraint \eqref{e59} in the definition of the ghost field $\varphi_{1}$ of \cref{e35}, one finds that:
\begin{equation}
\varphi_{1}\rightarrow -\dfrac{1}{\sqrt{\mu_{d}-\mu_{c}}}(\mu_{c}-m^{2}\epsilon^{2})\phi=O(\epsilon^{4}),
\end{equation}
where the definition \eqref{muc} of $\mu_{c}$ has been used in the last equality. On the other hand, the physical state $\varphi_{2}$ of \cref{e36} is now expressed as:
\begin{equation}
\varphi_{2}\rightarrow \dfrac{1}{\sqrt{\mu_{d}-\mu_{c}}}(\mu_{d}-m^{2}\epsilon^{2})\phi=(1-m^{2}\epsilon^{2})\phi +O(\epsilon^{4}).
\end{equation}
Treating the Lagrangian in the EFT context then naturally eliminates the ghost state $\varphi_{1}$ while keeping $\varphi_2$ in the effective theory. The Lagrangian \eqref{e37} indeed reads:
\begin{equation}
\mathcal{L}=\frac{1}{2}\varphi_{2}(\Box-\mu_{c}M^{2})\varphi_{2} +O(\epsilon^4) = \frac{1}{2}\varphi_{2}\Box \varphi_{2} -m^{2}(1+m^{2}\epsilon^{2})\varphi_{2}^2 +O(\epsilon^4),
\end{equation}
which exactly coincides with \eqref{e60}. We see that the advantage of the analysis made directly at the EFT level is to avoid the introduction of a new degree of freedom and the diagonalisation of the kinetic and mass terms. Field redefinitions or application of the JLM procedure take care of this treatment automatically by relegating the ghost modes to higher orders in the perturbative $\epsilon$ expansion.

\section{Higher-derivative operators in supersymmetric theories}\label{sec:HDSUSY}

In this section, we show how to generalise the ideas described above to the treatment of higher-derivative supersymmetric chiral Lagrangians. Higher-derivative terms for chiral superfields induce higher-derivative terms for the scalar component and derivative terms for the auxiliary field. These are  related by supersymmetry. We show how to take care of the extra degrees of freedom in the context of SUSY EFTs. 

A standard procedure for higher-derivative theories rewrites them as two-derivative theories containing additional (ghost) fields. When heavy, such fields can then be integrated out and a well-behaved low-energy Lagrangian is obtained. This path was followed in HD SUSY theories in \cite{Antoniadis:2007xc} and we come back to it in \cref{comparisonwithadditionalfields}.  The approach of the previous section, based on the equations of motion, avoids the introduction of new fields and the identification of ghosts and treats all HD theories in a systematic way.

We start this section by reviewing the general structure of the SUSY theories under consideration before introducing the HD SUSY operators relevant for this work. They are four space-time derivative operators, containing two, three or four chiral fields.  We then apply the methods described above to eliminate them in the EFT context, and show how this modifies the effective Lagrangian. We follow the SUSY conventions of \cite{Wess:1992cp}, recalled in \cref{AppendixSuperfields}.

\subsection{Global SUSY Lagrangians for chiral superfields with higher-derivative terms }

Effective field theories  include all operators compatible with the symmetries of the theory, leading generally to an infinite series of higher-derivative operators. We start our discussion by considering the generic supersymmetric chiral Lagrangian
\begin{align}
\mathcal{L}=&\int \!\! d^{4}\theta K (\Phi,\bar{\Phi}, D_{A}\Phi,D_{B}\bar{\Phi}, D_{A}D_{B}\Phi, \dots)+\left[\int\!\!d^{2}\theta\,W(\Phi,\partial_{\mu}\Phi,\bar{D}^{2}\bar{\Phi},\partial_{\mu}\partial_{\nu}\Phi,\dots)+h.c.\right].\label{e3.14}
\end{align}
The K\"ahler potential $K$ is real and includes an arbitrary number of superspace derivatives of $\Phi$ and $\bar{\Phi}$. The superpotential $W$ is a holomorphic function, including only chiral superfields. It can thus depend on $\Phi$ and its spacetime derivatives, and {\it a priori} also on $\bar{D}^{2}\bar{\Phi}$ and its spacetime derivatives. As noted in \cite{Ciupke:2016agp}, since the superspace derivatives $D_{A}$, introduced in \cref{AppendixSuperfields}, are the only objects that anti-commute with the supersymmetry generators, they are the only required ingredients to study higher-derivatives terms in global SUSY. 

We will consider SUSY Lagrangians expanded in inverse powers of a cutoff scale $M=1/\epsilon$, and irrelevant HD operators will thus be multiplied by powers $\epsilon$.

We study the HD corrections of $K$ and $W$ following \cite{Ciupke:2016agp}. We however do not use the superfield condition $\partial_{\mu}\Phi=\partial_{\mu}\bar{\Phi}=0$, imposed to reject terms not contributing to the scalar potential. Using \cref{stDvsSUSYD} and the chirality of $\Phi$, the K\"ahler potential introduced in \cref{e3.14} can be rewritten in terms of space-time derivatives and at most two SUSY covariant derivatives: \begin{align}
K= K(&\partial^{(i)}\Phi,\partial^{(i)}\bar\Phi, D^{\alpha}\partial^{(k)}\Phi ,\bar D_{\dot{\alpha}} \partial^{(k)}\bar \Phi, D^{2}\partial^{(p)}\Phi, \bar D^{2}\partial^{(p)}\bar \Phi), \label{Khigher}
\end{align}
where the operators $\partial^{(i)}\Phi, D_{\alpha}\partial^{(k)}\Phi, D^{2}\partial^{(p)}\Phi,$ have mass dimensions $i+1, k+\frac{3}{2}$ and $p+2$ respectively. In the above expression one should read $\partial^{(0)}\Phi\equiv\Phi$ and properly contract both Lorentz and spin indices.  To go from the Lagrangian \eqref{e3.14} with SUSY covariant derivatives to \eqref{Khigher}, with explicit space-time derivatives, one makes use of expressions such as \cref{boxPhi} or yet the slightly more involved:
\begin{equation}
-\frac{i}{64}(\bar{\sigma}^{\mu})^{\dot{\alpha}\beta} D_{\alpha}\bar{D}_{\dot{\alpha}} D_{\beta} \bar{D}^2D^2\Phi \propto D_{\alpha}\partial^{(3)} \Phi.
\end{equation}

The superpotential's dependence in higher-derivative operators turns out to be simpler once written as a finite series of chiral operators. As explained above, it could a priori  depend on the chiral superfields $\Phi$, $\bar{D}^{2}\bar{\Phi}$, and their derivatives.  However, terms containing $\bar{D}^{2}\bar{\Phi}$ and derivatives can be integrated by parts and included in the K\"ahler potential instead, using the identity of \cref{ibpsupefield}. Hence, we write the corrected superpotential as:
\begin{equation}
W=W(\partial^{(i)}\Phi).\label{Whigher}
\end{equation} 
There are some ambiguous higher-derivative terms such as:
\begin{equation}
\int d^2 \theta (\bar D^2\bar \Phi)^n + c.c. = -4 \int d^4 \theta \Big(\bar \Phi (\bar D^2\bar \Phi)^{n-1} + c.c.\Big). 
\end{equation}
According to the remark above, it should be seen as correction to the K\"ahler potential. As it only contains the chiral superfield $\bar D^2 \bar \Phi$ , it is however tempting to interpret it as a correction to the superpotential.
This shows that in SUSY theories with higher-derivative operators, the distinction between K\"ahler potential and superpotential is no longer unique. This was to be expected since these functions were introduced for two-derivative SUSY theories.

Let us comment on the use of SUSY derivatives or space-time derivatives. Higher-derivative operators refer to space-time derivatives, hence \cref{Khigher,Whigher} have been written in terms of higher space-time derivative operators. In this way, the link with the field theory discussion of previous section is clear. One can however express all the higher-order space-time derivatives as properly-contracted anti-commutators of SUSY covariant derivatives, using \cref{stDvsSUSYD} repeatedly. Superspace calculus and integration being facilitated by the use of SUSY derivatives only, in the following we will construct specific Lagrangians using only the latter.

In what follows  we will consider a SUSY theory with irrelevant HD operators appearing in the action with the adequate power of the EFT expansion parameter $\epsilon=1/M$.
As marginal part, we will only consider the canonical K\"ahler potential of 4d chiral SUSY Lagrangians:
\begin{equation}
K_0(\Phi,\bar{\Phi})=\Phi \bar{\Phi} \,. \label{K0}
\end{equation}
This choice does not change the discussion on HD terms and one could start from other K\"ahler potentials. As irrelevant part, we will mainly focus on operators with four space-time derivatives. A subset of these four-derivative operators, like  the supersymmetric extension of the $P(X,\phi)$, leads to second-derivative equations of motion and so does not introduce additional (ghost) degrees of freedom. Other four-derivative operators could lead to the presence of ghosts. As noted in \cite{Antoniadis:2007xc,Khoury:2010gb,Ciupke:2015msa,Bielleman:2016grv} such higher-derivative operators will in general also introduce derivative terms for the auxiliary field of the chiral multiplet. In the past such operators were either discarded \cite{Khoury:2010gb,Ciupke:2015msa} or treated by setting the derivatives of the auxiliary field to zero by hand \cite{Bielleman:2016olv}.  In \cite{Antoniadis:2007xc} some of these operators were carefully studied by introducing new dynamical fields. In this way, the authors showed how to bring the original theory into a two-derivative one with additional ghost fields.  We will explain how to treat all possible HD operators at a certain order of $\epsilon$ expansion in a systematic way, by consistently treating these terms in an EFT approach.

We essentially focus on theories which do not correct the superpotential, with the caveat explained below \cref{Whigher}. Unless otherwise stated, we will thus consider the uncorrected superpotential:
\begin{equation}
W=W(\Phi).
\end{equation}

\subsection{Four-derivative corrections to the K\"ahler potential}

We consider operators with two additional spacetime derivatives with respect to the canonical kinetic term. As can be seen in \cref{K0}, the latter is written as an integral over the whole superspace of an operator without covariant derivatives. According to \cref{stDvsSUSYD}, for a generic chiral superfield $\Phi$ we have:
\begin{equation}
\bar{D} D \Phi \sim \partial^{\mu} \Phi,
\end{equation}
so that four-derivative operators correcting the K\"ahler potential shall be constructed with four SUSY covariant derivatives. Most of the operators we consider below were studied in the past in various contexts \cite{Antoniadis:2007xc,Antoniadis:2009rn,Dudas:2015vka,Delgado:2023ivp}.

\paragraph{Two chiral superfields} The only four-covariant-derivative operators correcting the K\"ahler potential including two fields read:
\begin{equation}
\mathcal{N}_1 = \bar\Phi \bar D^2 D^2 \Phi, \qquad \mathcal{N}'_1 = \bar D^2 \bar\Phi  D^2 \Phi. \label{2fields4deriv}
\end{equation}
Once integrated over the whole superspace to form a Lagrangian, they can be related by partial integration. We can thus consider only the first operator, with top bosonic component:
\begin{equation}
  \frac{1}{16} \int d^4\theta \mathcal{N}_1 = \Box\phi \Box \bar \phi + F\Box \bar F \, .
\end{equation}

\paragraph{Three chiral superfields} The four-derivative operators correcting the K\"ahler potential with two chiral and one anti-chiral fields are:
\begin{align}
&\mathcal{M}_1= \Phi D^2\Phi \bar D^2 \bar\Phi,  \qquad && \mathcal{M}'_1=  \Phi \bar{\Phi}  \bar D^2 D^2 \Phi,  \label{3fields4deriv1}\\
&\mathcal{M}_2=\bar D^2 \bar\Phi D^{\alpha}\Phi D_{\alpha}\Phi, \qquad && \mathcal{M}'_2= \Phi \bar D D\Phi D\bar D\bar \Phi \, . \label{3fields4deriv2}
\end{align}
To construct real corrections to the K\"ahler potential, such operators should be accompanied by their hermitian conjugates. Note that  integrating by parts in superspace and using \cref{superderivformula} shows that:
\begin{equation}
\mathcal{M}'_1= \mathcal{M}_1 + {\rm tot.}\, , \qquad  2\mathcal{M}'_2 = \mathcal{M}_1+\mathcal{M}_2 + {\rm tot.}\, ,
\end{equation}
where {\rm tot.} denote total derivatives. Hence, only two operators (and their conjugates) give distinct corrections to the K\"ahler potential at this level. Their top bosonic components read:
\begin{align}
   & \frac{1}{16}\int d^4\theta \mathcal{M}_1= \phi \Box \phi \Box \bar\phi + F \bar F \Box \phi + \phi \bar F\Box F + \rm{tot.} \, , \\
   &\frac{1}{16}\int d^4\theta \mathcal{M}_2=\p_{\mu}\phi\p^{\mu}\phi\Box\bar{\phi}+2\bar F \p_{\mu}\phi\p^{\mu}F + \rm{tot.} \, .
\end{align}

\paragraph{Four chiral superfields}Several different operators can be assembled with four covariant derivatives, two chiral and two anti-chiral superfields \cite{Khoury:2010gb}:
\begin{align}
\mathcal{O}_{0}&=D\Phi D\Phi\bar{D}\bar{\Phi}\bar{D}\bar{\Phi}, \quad &&\mathcal{O}_{1}=|\Phi|^{2}D^{2}\Phi\bar{D}^{2}\bar{\Phi}, \quad &&&\mathcal{O}_{2}=\Phi D^{2}\Phi\left(\bar{D}\bar{\Phi}\right)^{2}, \label{4fields4deriv}\\
\mathcal{O}_{3}&=|\Phi|^{2}D\bar{D}\bar{\Phi}\bar{D}D\Phi, \quad &&\mathcal{O}_{4}=\Phi^{2}D\bar{D}\bar{\Phi}D\bar{D}\bar{\Phi}, \quad &&& \mathcal{O}_{5}=\Phi D\Phi\bar{D}\bar{\Phi}D\bar{D}\bar{\Phi}. 
\end{align}
One should also consider additional operators made from complex conjugation of the above operators, if not already in the list. As in the case with fewer fields, some operators are redundant and can be related to others via superspace integration by parts. For instance, we have that:
\begin{equation}
2\mathcal{O}_5=\mathcal{O}_0+\mathcal{O}_2 + {\rm tot.}\, .
\end{equation}
In fact, it is enough to consider only the first line of the list, namely operators constructed from $\mathcal{O}_0$, $\mathcal{O}_1$, and $\mathcal{O}_2$ and their complex conjugates.
Their top bosonic components read, up to total derivatives:
\begin{align}
& \frac{1}{16} \int d^4 \theta \, {\mathcal{O}_{0}}=\mathrm{A}-2\mathrm{B}+\mathrm{C}\label{e3.36},\\
&\frac{1}{8} \int d^4 \theta \, {\mathcal{O}_{1}} =2\mathrm{C}+2\mathrm{D}+\mathrm{E}+3\mathrm{F}+2\mathrm{G}\label{e3.37},\\
& \frac{1}{16}\int d^4\theta\, (\mathcal{O}_{2} + {\rm c.c.}) =-2\mathrm{C}-2\mathrm{F}-2\mathrm{G}+\mathrm{H}\label{e3.39},
\end{align}
where we defined the following component expressions:
\begin{align}
\mathrm{A}\equiv&(\partial \phi)^{2}(\partial \bar \phi)^{2}, \quad &&\mathrm{B}\equiv|F|^{2}|\partial \phi|^{2}\label{e3.29},\\
\mathrm{C}\equiv&|F|^{4}, \quad &&\mathrm{D}\equiv|\phi|^{2}|\Box \phi|^{2}\label{e3.31},\\
\mathrm{E}\equiv&|\phi|^{2}F\Box \bar F +|\phi|^{2}\bar F \Box F, \quad &&\mathrm{F}\equiv|F|^{2} \phi \Box \bar \phi+|F|^{2}\bar \phi\Box \phi\label{e3.33},\\
\mathrm{G}\equiv&\phi \bar F \partial \bar \phi\cdot\partial F+\bar \phi F\partial \phi \cdot\partial \bar F , \quad &&\mathrm{H}\equiv \phi \Box \phi (\partial \bar \phi)^{2}+\bar \phi\Box \bar \phi(\partial \phi )^{2}. \label{e3.35}
\end{align}
We use the notation $|\partial \phi |^{2}\equiv\partial_{\mu} \phi \partial^{\mu} \bar \phi$. The operator  $\mathcal{O}_0$ is special since it contains no derivative of $F$ and its bosonic part only contains a top component. 

For completeness, we mention that in addition to the operators considered above, one could also consider operators with three chiral and one anti-chiral superfields (and their complex conjugates). There are two independent operators \cite{Khoury:2010gb}:
\begin{equation}
\mathcal{P}_1=\Phi^2D^2\Phi\bar D^2 \bar\Phi, \qquad \mathcal{P}_2=\Phi D\Phi D\Phi \bar D^2\bar\Phi.
\end{equation}

\subsection{Treatment of  higher-derivative Lagrangians in components}

In this section, we show how to study a particular SUSY Lagrangian including all possible four-derivative terms with two chiral and two anti-chiral superfields, seen as the lowest-order expansion of a full theory.
In this section, we work in component form, postponing the superspace study to \cref{superspaceHDtreatment}.

We consider a theory with an uncorrected superpotential $W$. The K\"ahler potential $K_0$ of \cref{K0} is corrected by the irrelevant operators of \cref{e3.36,e3.37,e3.39}. These all have mass dimension eight and therefore appear in the Lagrangian multiplied by $1/M^{4}$. We then choose the expansion parameter to be $\epsilon=1/M$, and parameterise the Lagrangian in terms of dimensionless constants $\alpha$, $\beta$, $\gamma$, as:
\begin{align}
\mathcal{L}=\mathcal{L}_0+\epsilon^2 \mathcal{L}_{HD}=&\int d^4\theta \left\{ K_{0}(\Phi,\bar \Phi) + \frac{\epsilon^4}{16} \Big( \frac{\alpha}{2} \mathcal{O}_0 -\beta {\mathcal{O}}_2 + \gamma ({\mathcal{O}}_1+{\mathcal{O}}_2) +{ \rm c.c.} \Big) \right\} \nonumber\\
&+ \int d^2 \theta W(\Phi)  + O(\epsilon^5) . \label{lagrangianSuperspace1}
\end{align}
According to \cref{e3.39,e3.33,e3.35}, the Lagrangian in components thus reads:
\begin{align}
\mathcal{L} =&-\partial^{\mu} \phi \p_{\mu} \bar \phi+|F|^{2}+\frac{\partial W}{\partial \phi}F+\frac{\partial \bar{W}}{\partial \bar \phi}\bar F \nonumber\\
&+\epsilon^{4}\left\{\alpha(\partial \phi)^{2}(\partial \bar \phi)^{2}+2\gamma|\phi|^{2}|\Box \phi|^{2} -\left(\beta-\gamma\right)\left(\phi \Box \phi(\partial \bar \phi)^{2}+\bar \phi\Box \bar \phi(\partial \phi)^{2}\right)\right. \nonumber\\
&+\left(\alpha+2\beta\right)|F|^{4}-2\alpha|F|^{2}|\partial \phi|^{2}+\left(2\beta+\gamma\right)\left(|F|^{2}\phi \Box \bar \phi+|F|^{2}\bar \phi\Box \phi\right)\nonumber\\
& \left. + \gamma\left(|\phi|^{2}F\Box \bar F +|\phi|^{2}\bar F \Box F\right)+2\beta\left(\phi \bar F \partial \bar \phi\cdot\partial F+\bar \phi F\partial \phi\cdot\partial \bar F \right) \right\}+\mathcal{O}(\epsilon^5)\ . 
\label{e3.50}
\end{align}

As motivated above, we view this Lagrangian as an EFT valid to order $O(\epsilon^{5})$ of an unknown but well-defined UV theory. In this sense, we assume that the presence of ghost modes is an artefact of the truncation below the energy scale $M$. Treating this Lagrangian with the methods reviewed in \cref{JLMfields} allows for the rewriting of the theory in terms of an equivalent Lagrangian, valid below the energy scale $M$, containing the same dynamical degrees of freedom as the $\mathcal{O}(\epsilon^0)$ theory and where the ghosts are absent.

\subsubsection{Eliminating derivatives of the auxiliary field $F$}

We turn our attention to $F$, the auxiliary field of two-derivative chiral SUSY theories. In the $O(\epsilon^{4})$ higher-derivative perturbation under study, derivatives $\partial F$ or $\Box F$ appear and can lead to questions regarding the nature of $F$. It could possibly become dynamic, signalling the appearance of new degrees of freedom. This can be understood in the SUSY context from the fact that higher-derivative terms give rise to additional degrees of freedom for the dynamical field $\phi$. The new degrees of freedom of $\phi$ and $F$ are expected to be related by supersymmetry \cite{Dudas:2015vka}. In the EFT context, we thus expect that the would-be dynamical part of $F$ can be eliminated in the low-energy theory, in the same manner as the ghost degrees of freedom associated to the higher derivatives of $\phi$. The field $F$ should thus stay non-dynamical in  the low-energy theory, after elimination of its derivative terms.  We now show that this elimination is indeed possible by performing field redefinitions or equivalently by  using the constraints of the JLM procedure.

\paragraph{Field redefinitions} We now show how to eliminate the terms containing derivatives of the auxiliary field $F$ in the Lagrangian \eqref{e3.50}. For this purpose, it is sufficient to study the $F$-dependent part of the Lagrangian:
\begin{align}
\mathcal{L}_F+ \mathcal{L}_{\partial F} & = |F|^{2}+\frac{\partial W}{\partial \phi}F+\frac{\partial \bar{W}}{\partial \bar \phi}\bar F  +  \epsilon^4 (\alpha+2\beta) |F|^4 \nonumber\\
&\quad + \epsilon^{4} \left(\bar  \Upsilon \cdot\partial F+ \Upsilon\cdot\partial \bar F + \bar T\Box F+T\Box \bar F \right)+\mathcal{O}(\epsilon^5). \label{termsFLagrangian}
\end{align}
The part $\mathcal{L}_{\partial F}$, containing derivatives of $F$, is defined as the second line of \cref{termsFLagrangian}. It is expressed in terms of the functions:
\begin{equation}
T\equiv \gamma F |\phi|^{2}, \qquad  \Upsilon_{\mu}\equiv 2\beta F \bar \phi\partial_{\mu}\phi\  .
\end{equation} 
The term  $\mathcal{L}_ F$, without derivatives of $F$, is defined as the first line of  \cref{termsFLagrangian} and contains the usual algebraic terms in $F,\bar{F}$, supplemented with an $|F|^4$ arising from the HD operators.

 We can now transform the Lagrangian \eqref{termsFLagrangian} to eliminate terms containing derivatives of $F$, making use of the field redefinition:
\begin{equation}
F\longrightarrow F+\epsilon^{4}\left(\partial\cdot \Upsilon-\Box T\right),  \label{fieldredefF}
\end{equation}
and the similar one for $\bar F $. We used the notation $ \partial\cdot \Upsilon\equiv\partial^{\mu}\Upsilon_{\mu}$. Under this redefinition, the Lagrangian transforms  as:
\begin{align}
\mathcal{L}_F+\mathcal{L}_{\partial F} \longrightarrow \,\, &\mathcal{L}_F+\mathcal{L}_{\partial F}+\epsilon^{4}\left(\partial\cdot \Upsilon-\Box T\right) \frac{\partial\mathcal{L}_{0}}{\partial F}+\epsilon^{4}\left(\partial\cdot \Upsilon^*-\Box T^*\right)\frac{\partial\mathcal{L}_{0}}{\partial \bar F }+O(\epsilon^{5})\nonumber\\
\!=& \,\mathcal{L}_F+\mathcal{L}_{\partial F}+ \epsilon^{4}\left(\partial\cdot \Upsilon-\Box T\right)  \!\left[\bar F +\frac{\partial W}{\partial \phi}\right] \! + \epsilon^{4}\left(\partial\cdot \Upsilon^*-\Box T^*\right)  \! \left[F+\frac{\partial \bar{W}}{\partial \bar \phi}\right] \! +O(\epsilon^{5})\nonumber\\
=&\, \mathcal{L}_{F}-\epsilon^{4}\left( \Upsilon\cdot\partial\left( \frac{\partial W}{\partial \phi}\right) + T\Box \frac{\partial W}{\partial \phi }\right)- \epsilon^{4}\left(\bar \Upsilon  \cdot\partial \left(\frac{\partial \bar{W}}{\partial \bar \phi}\right)+  \bar T   \Box \frac{\partial \bar{W}}{\partial \bar \phi}\right) + O(\epsilon^5) ,\nonumber \\
\label{n2.55}
\end{align}
where we have used integration by parts in the last step. We comment now on the fact that the full Lagrangian \eqref{e3.50} presents additional terms containing the auxiliary field $F$, at order $O(\epsilon^2)$, appearing without derivatives. Under the field redefinition \eqref{fieldredefF} these terms only generate $O(\epsilon^4)$ terms. We thus have shown using the above field redefinition that the full Lagrangian of \eqref{e3.50} is equivalent to:
\begin{align}
 \mathcal{L} =&-|\partial \phi|^{2}+|F|^{2}+\frac{\partial W}{\partial \phi}F+\frac{\partial \bar{W}}{\partial \bar \phi}\bar F \nonumber\\
 &+\epsilon^{4}\left\{ \vphantom{\Big(^2}\alpha(\partial \phi)^{2}(\partial \bar \phi)^{2}+2\gamma|\phi|^{2}|\Box \phi |^{2} -\left(\beta-\gamma\right)\left(\phi \Box \phi(\partial \bar \phi)^{2}+\bar \phi\Box \bar \phi(\partial \phi)^{2}\right)\right. \nonumber\\
 &+\left(\alpha+2\beta\right)|F|^{4}-2\alpha|F|^{2}|\partial \phi|^{2}+\left(2\beta+\gamma\right)\left(|F|^{2}\phi \Box \bar \phi+|F|^{2}\bar \phi\Box \phi \right)\nonumber\\
 & \left. - \Upsilon\cdot\partial\left( \frac{\partial W}{\partial \phi}\right) - T\Box \frac{\partial W}{\partial \phi} - \bar \Upsilon  \cdot\partial \left(\frac{\partial \bar{W}}{\partial \bar \phi}\right) -  \bar T   \Box \frac{\partial \bar{W}}{\partial \bar \phi}  \right\} + O(\epsilon^5)\ . \label{LwithoutdF}
\end{align}

\paragraph{JLM constraints for $F$} When equivalent to field redefinitions, plugging the JLM constraints in the Lagrangian gives a systematic way to treat the $F$ derivatives. The equation of motion for $\bar F $ is:
\begin{equation}
\frac{\partial\mathcal{L}}{\partial \bar F }-\partial_{\mu}\frac{\partial\mathcal{L}}{\partial (\partial_{\mu}\bar F )}+\partial_{\mu}\partial_{\nu}\frac{\partial\mathcal{L}}{\partial (\partial_{\mu}\partial_{\nu}\bar F )}=0,
\end{equation}
and when applied to the Lagrangian of \eqref{e3.50}, it takes the explicit form:
\begin{align}
F+\frac{\partial \bar{W}}{\partial \bar \phi}+\epsilon^{4}\big\{&(-2\alpha-2\beta+2\gamma)F|\partial \phi |^{2}+2(\alpha+2\beta)F|F|^{2}\nonumber\\
&+2\gamma|\phi|^{2}\Box F+2\gamma \bar \phi F\Box \phi+(2\beta+2\gamma)\phi F\Box \bar \phi\nonumber\\
&+(2\beta+2\gamma) \phi\partial \bar \phi\cdot\partial F-(2\beta-2\gamma) \bar \phi\partial \phi\cdot\partial F\big\}=O(\epsilon^{5})\label{eomFstar}\,\,.
\end{align}
Strictly following the JLM procedure described in \cref{JLMfields}, the first constraints for $F$ are obtained by multiplying \eqref{eomFstar} by the expansion parameter $\epsilon$, getting the trivial expression:
\begin{align}
\epsilon&\left[F+\frac{\partial \bar W}{\partial \bar \phi } \right]=O(\epsilon^{5})\label{e3.58}\,\,.
\end{align}
The full set of constraints is then obtained by multiplying again by $\epsilon^3$ and differentiating, leading to: 
\begin{align}
\epsilon^{4}&\left[F+\frac{\partial \bar W}{\partial \bar \phi }\right]=O(\epsilon^{8})\label{e3.60}\, , \\
\epsilon^{4}&\left[\partial_{\mu}F+\frac{\partial^{2} \bar W}{\partial \bar \phi^{2}} \partial_{\mu}\bar \phi\right]=O(\epsilon^{8})\label{e3.62}\,, \\
\epsilon^{4}&\left[\Box F+\frac{\partial^{3} \bar W}{\partial \bar \phi^{3}} (\partial \bar \phi)^{2}+\frac{\partial^{2} \bar W}{\partial \bar \phi^{2}} \Box \bar \phi\right]=O(\epsilon^{8})\label{e3.64}\,\,,
\end{align}
with similar expressions holding for $\bar F $. We can then plug these last two equations in \eqref{eomFstar} to get the algebraic equation of motion for $F$:
\begin{align}
F+\frac{\partial \bar W}{\partial \bar \phi}+\epsilon^{4}\bigg\{&(-2\alpha-2\beta+2\gamma)F|\partial \phi|^{2}+2(\alpha+2\beta)F|F|^{2}+2\gamma \bar \phi F\Box \phi+(2\beta+2\gamma) \phi F\Box \bar \phi\nonumber\\
&-2\gamma|\phi|^{2}\left[\frac{\partial^{3} \bar W}{\partial \bar \phi^{3}}(\partial \bar \phi)^{2}+\frac{\partial^{2} \bar W}{\partial \bar \phi^{2}}\Box \bar \phi\right]\nonumber\\
&-(2\beta+2\gamma) \phi \frac{\partial^{2} \bar W}{\partial \bar \phi^{2}} (\partial \bar \phi)^{2}+(2\beta-2\gamma) \bar \phi \frac{\partial^{2} \bar W}{\partial \bar \phi^{2}} |\partial \phi|^{2}\bigg\}=O(\epsilon^{5})\label{e3.70}\,\,.
\end{align}
When used directly in the Lagrangian \eqref{e3.50}, the above constraints give exactly the Lagrangian \eqref{n2.55} obtained by field redefinitions and hence lead to the same equation of motion of \eqref{e3.70}. The equivalence between the two methods is then clear at the order we are working on. This is in agreement with the comments below \cref{e49} on the use of the JLM constraints in the Lagrangian, since we are treating $\epsilon^4$ corrections in an approximation at order $O(\epsilon^5)$.

\subsubsection{Integrating out the auxiliary field}
 
At this stage, we are left with the Lagrangian \eqref{LwithoutdF} which contains no derivatives of $F$. The latter is thus indubitably a non-propagating auxiliary field. The Lagrangian takes the form:
\begin{equation}
\mathcal{L}\supset  (1+\epsilon^4 f)|F|^{2}+\left(\frac{\partial W}{\partial \phi}+\epsilon^4 g\right)F+\left(\frac{\partial \bar{W}}{\partial \bar{\phi}}+\epsilon^4 \bar{g}\right)\bar F +\epsilon^4 k|F|^{4}\,\,,
\end{equation}
with $k=\alpha+2\beta$ constant and $f$ and $g$ functions of $\phi$, $\bar \phi$, and their derivatives, that can be read from \cref{LwithoutdF}.

The equation of motion \eqref{e3.70} is solved perturbatively as:
\begin{equation}
F=F_0+\epsilon^4 F_4 + O(\epsilon^5), \qquad F_0 \equiv -\frac{\partial \bar W}{\partial \bar \phi}\ .
\end{equation}
In the $O(\epsilon^4)$ part of the Lagrangian, $F$ can be replaced by $F_0$. On the other hand, the zeroth order part could in principle generate new $O(\epsilon^4)$ terms coming from $\epsilon^4 F_4$. Nevertheless, we see that the part of the zeroth order Lagrangian depending on the auxiliary field reads
\begin{align}
|F|^{2}+\frac{\partial W}{\partial \phi} F+\frac{\partial \bar W}{\partial \bar \phi} \bar F &=|F_0|^2+\left( \frac{\partial W}{\partial \phi} F_0 +  \epsilon^4 F_4 \Big(F_0^*+\frac{\partial W}{\partial \phi}\Big) + c.c. \right)+ O(\epsilon^5)\nonumber\\
&=-\left|\frac{\partial W}{\partial \phi} \right|^2 +O(\epsilon^5).
\end{align}
Hence, the scalar potential obtained from the zeroth order Lagrangian computed with the corrected auxiliary field is the same, at order $O(\epsilon^5)$, as the one computed from the uncorrected one. We denote it:
\begin{equation}
V(\phi,\bar \phi) \equiv\bigg|\frac{\partial W}{\partial \phi}\bigg|^{2}=|F_0|^2,
\end{equation}
and recall that it is corrected by HD terms, as we see below. To conclude, the zeroth order value $F_0$ is sufficient to evaluate the full Lagrangian at order $O(\epsilon^5)$.

\subsubsection{Higher-derivative scalar field Lagrangian and final physical Lagrangian} 

After integrating out the auxiliary field $F$ as described in the previous subsection, we obtain the Lagrangian for the scalar $\phi$:
\begin{align}
\mathcal{L}=&-|\partial \phi|^{2}-V+\epsilon^{4}\big\{\alpha(\partial \phi)^{2}(\partial \bar \phi)^{2}-2\alpha V|\partial \phi|^{2}\nonumber\\
	       &+\left(\alpha+2\beta\right)V^{2}+2\gamma|\phi|^{2}|\Box \phi |^{2}\nonumber\\
		   &+\gamma|\phi|^{2}\left(V_{\phi}\Box \phi+V_{\phi\phi}(\partial \phi)^{2}+V_{\bar \phi}\Box \bar \phi+V_{\bar \phi\bar \phi}(\partial \bar \phi)^{2}\right)\nonumber\\
		   &+\left(2\beta+\gamma\right)V\left(\phi \Box \bar \phi+\bar \phi\Box \phi \right)+2\beta\left(\phi V_{\bar \phi}(\partial \bar \phi)^{2}+\bar \phi V_{\phi}(\partial \phi)^{2}\right)\nonumber\\
		   &-\left(\beta-\gamma\right)\left(\phi\Box \phi(\partial \bar \phi)^{2}+\bar \phi\Box \bar \phi(\partial \phi)^{2}\right)\big\}+O(\epsilon^{5}) \, ,\label{e3.76}
\end{align} 
where the subscripts denote partial derivatives with respect to the scalars, e.g. $V_\phi\equiv{\partial V }/{\partial \phi}$. 
The equation of motion for $\phi$ reads:
\begin{align}
\Box \phi-V_{\bar \phi}+\epsilon^{4} \cdots =O(\epsilon^{5}),
\end{align}
so that the first constraints of the JLM procedure are:
\begin{align}
&\epsilon\left[\Box \phi-V_{\bar \phi}\right]=O(\epsilon^{5})\label{e3.88}\,\,,
\end{align}
and its complex conjugate, together with their derivatives $\partial^{(s)}$. The only constraints needed are the ones necessary to eliminate the $\epsilon^{4}\Box \phi$ terms in \eqref{e3.76}. They are obtained by multiplying \eqref{e3.88} by $\epsilon$, which is the second order, $k=n-1$, of the JLM procedure. At the order $O(\epsilon^5)$ under consideration, the JLM constraints can be used directly in the Lagrangian, leading to: 
\begin{align}
\mathcal{L}=&-|\partial \phi|^{2}-V+\epsilon^{4}\big\{\alpha(\partial \phi)^{2}(\partial \bar \phi)^{2}-2\alpha V|\partial \phi|^{2}\nonumber\\
		   &+\left(\gamma|\phi|^{2}V_{\phi\phi}+(\beta+\gamma)\bar \phi V_{\phi}\right)(\partial \phi)^{2}\nonumber\\
		   &+\left(\gamma|\phi|^{2}V_{\bar \phi\bar \phi}+(\beta+\gamma)\phi V_{\bar \phi}\right)(\partial \bar \phi)^{2}\nonumber\\
		   &+\left(\alpha+2\beta\right)V^{2}+4\gamma|\phi|^{2}|V_{\phi}|^{2}+\left(2\beta+\gamma\right)V\left(\phi V_{\phi}+\bar \phi V_{\bar \phi}\right)\big\}+O(\epsilon^{5})\,\,,\label{e3.89}   
\end{align}
which only depends on $\phi$, $\bar \phi$, and their first derivatives, and is on-shell equivalent to the initial Lagrangian.

Upon integrating by parts, the kinetic terms of the above Lagrangian can be brought to a standard form, including only $\p_{\mu} \phi \p^{\mu} \bar \phi$. For instance, we can write:
\begin{align}
\epsilon^4 \bar \phi V_{\phi} (\partial \phi)^2&= \epsilon^4  \bar \phi W'' \bar{W}' \p_{\mu}\phi \p^{\mu}\phi=\epsilon^4  \bar \phi \bar{W}' \p_{\mu}W' \p^{\mu}\phi \nonumber\\
&=- \epsilon^4 (|W'|^2+ \bar \phi W' \bar{W}'' )\p_{\mu}\bar \phi\p^{\mu}\phi -\epsilon^4 \bar \phi |W'|^2 \bar{W}'' W' + \,\, {\rm tot.} \nonumber\\
&=-\epsilon^4(V + \bar \phi V_{\bar \phi})\p_{\mu}\bar \phi\p^{\mu}\phi - \epsilon^4  \bar \phi V V_{\bar \phi} + \,\, {\rm tot.}  \, .
\end{align}
In the second line, we used once again the constraint on $\epsilon^4 \Box \phi$ derived from \cref{e3.88}. Similarly, one can write the following combination as:
\begin{align}
 \epsilon^4 (|\phi|^2 V_{\phi\phi} + \bar \phi V_{\phi}) (\p \phi)^2
 &=-\epsilon^4(\phi V_{\phi}+|\phi|^2 V_{\phi \bar \phi})\p_{\mu} \bar \phi \p^{\mu} \phi -\epsilon^4 |\phi|^2 |V_{\phi}|^2 + {\rm tot.}\, .
\end{align}
Hence, the second and third lines of \cref{e3.89} can be written as:
\begin{align}
\mathcal{L}& \supset -  \epsilon^4  ( 2 \beta V + (\beta+\gamma)( \phi V_{\phi} +  \bar \phi V_{\bar \phi}) +  2 \gamma |\phi|^2 V_{\phi\bar \phi})\p_{\mu}\bar \phi\p^{\mu}\phi \nonumber\\
& \quad - \epsilon^4 (2 \gamma |\phi|^2 |V_{\phi}|^2 + \beta \bar \phi V V_{\bar \phi} +\beta  \phi V V_{\phi}).
\end{align}
The total Lagrangian, up to a total derivative, is equal to:
\begin{align}
\mathcal{L}=&-|\partial \phi|^{2}-V+\epsilon^{4} \alpha(\partial \phi)^{2}(\partial \bar \phi)^{2} \nonumber \\
&-\epsilon^4 \left\{2(\alpha+\beta) V + (\beta+\gamma)( \phi V_{\phi} +  \bar \phi V_{\bar \phi}) +  2 \gamma |\phi|^2 V_{\phi \bar \phi} \right\}|\partial \phi|^{2}\nonumber\\
		 &+\epsilon^{4}\left\{\left(\alpha+2\beta\right)V^{2}+2 \gamma|\phi|^{2}|V_{\phi}|^{2}+\left(\beta+\gamma\right)\left(\phi VV_{\phi}+\bar \phi VV_{\bar \phi}\right)\right\}+O(\epsilon^{5}), \label{finalLcomponents}
\end{align}
and it contains:
\begin{itemize}
\item[$\triangleright$] A non-standard kinetic term
	  \begin{equation}
	  -(1+\epsilon^4 f(\phi,\bar \phi))|\partial \phi|^{2}, \quad
	  f\equiv 2(\alpha+\beta) V + (\beta+\gamma)( \phi V_{\phi} +  \bar \phi V_{\bar \phi}) +  2 \gamma |\phi|^2 V_{\phi \bar \phi}.
	  \end{equation}
\item[$\triangleright$] A higher-derivative term:
	  \begin{equation}
	  \epsilon^{4}  \alpha (\partial \phi)^{2}(\partial \bar \phi)^{2}.
	  \end{equation}
\item[$\triangleright$] A modified scalar potential
	  \begin{align}
	 V_{tot}&= V(\phi ,\bar \phi)-\epsilon^{4}\big\{\left(\alpha+2\beta\right)V^{2}+\left(\beta+\gamma\right)\left(\phi VV_{\phi}+\bar \phi VV_{\bar \phi}\right)+2\gamma|\phi|^{2}|V_{\phi}|^{2}\big\}\nonumber\\
  &=V(\phi ,\bar \phi)- \epsilon^4 V \left(f(\phi ,\bar \phi) - \alpha V\right), \label{correctedVtot}
	  \end{align}
   where $V(\phi,\bar{\phi})=\partial_\phi W \partial_{\bar{\phi}} \bar{W}$ and we used the identity $V V_{\phi \bar \phi}=|V_{\phi}|^2$. 
\end{itemize}

Each ghostly higher-derivative term has been eliminated, leaving a Lagrangian with no unphysical degrees of freedom. One can obtain different physical effective theories by choosing different values of the three free coefficients $\alpha$, $\beta$ and $\gamma$.

Note that the correction to the scalar potential comes from the elimination of the higher-derivative terms through the redefinition of the field variables. In absence of a scalar potential, the corrections are absent. 

\subsection{Treatment of higher-derivative Lagrangians in superspace} \label{superspaceHDtreatment}
In this subsection, we show how the treatment of the HD Lagrangian under consideration can be performed entirely at the level of superfields, without using the component expansion. This renders the relation between the higher-derivatives of the auxiliary field and those for the scalar field explicit. We will see that the higher-derivatives are treated all at once. 
We recall that the superspace Lagrangian under study reads:
\begin{align}
\mathcal{L}=\mathcal{L}_0+\epsilon^4 \mathcal{L}_{HD}=&\int d^4\theta K_{0}(\Phi,\bar \Phi) + \int d^2 \theta W(\Phi) \nonumber\\
&+\epsilon^4 \int d^4\theta \frac{1}{16} \Big( \frac{\alpha}{2} \mathcal{O}_0 -\beta {\mathcal{O}}_2 + \gamma ({\mathcal{O}}_1+{\mathcal{O}}_2) +{ \rm c.c.} \Big)  + O(\epsilon^5) . \label{lagrangianSuperspace}
\end{align}

\subsubsection{Superfield equations of motion}

The first step of the JLM procedure is to write the equation of motion as an expansion in $\epsilon=1/M$ parameter.
Let us warm up with the standard field equation for a renormalisable chiral SUSY Lagrangian:
\begin{align}
\mathcal{L}&=\int d^4\theta K_0(\Phi,\bar{\Phi}) + \int d^2 \theta W(\Phi)=\int d^4\theta \, \Phi\bar{\Phi} + \int d^2 \, \theta W(\Phi)\nonumber\\
&=\int d^2\theta \left(-\frac{1}{4}\bar{D}^2 \bar{\Phi}\right)\Phi + W(\Phi),
\end{align}
from which we derive the standard 0-th order superfield equation of motion:
\begin{equation}
-\frac{1}{4}\bar{D}^2 \bar{\Phi} + \frac{\partial W}{\partial \Phi}(\Phi)=0. \label{0orderEOM}
\end{equation}
One can then treat higher-derivative operators in superspace in the same way to obtain the higher-order equation of motion. Varying the operators with respect to the superfield $\Phi$ gives:
\begin{align}
\int d^4\theta \,\, \delta \mathcal{O}_2 &=\int d^4\theta \,\,\delta(\Phi D^2 \Phi) \left(\bar{D}\bar{\Phi}\right)^2=\int d^4\theta \left( D^2 \Phi \left(\bar{D}\bar{\Phi}\right)^2+  D^2 \left(\Phi\left(\bar{D} \bar{\Phi}\right)^2\right)\right) \delta\Phi \nonumber\\
&= -\frac{1}{2}\int d^2\theta\bar{D}^2\left( D^2 \Phi \left(\bar{D}\bar{\Phi}\right)^2 + 2 D\Phi D\bar{D}\bar{\Phi}\bar{D}\bar{\Phi}\right)\delta \Phi, \\
\nonumber\\
\int d^4\theta \,\, \delta \mathcal{O}_1 &=\int d^4\theta \,\, \delta(\Phi D^2\Phi)\bar\Phi \bar D ^2 \bar\Phi =  \int d^4\theta \,\, \left( \vphantom{\left(\bar{\Phi}\right)^2} \bar\Phi D^2 \Phi \bar{D}^2\bar{\Phi}+  D^2 \left(\Phi \bar\Phi \bar{D}^2 \bar{\Phi}\right)\right) \delta\Phi \nonumber\\
&=-\frac{1}{4}\int d^2\theta \bar D^2 \left(\vphantom{\left(\bar{\Phi}\right)^2} 2 \bar\Phi D^2 \Phi \bar{D}^2\bar{\Phi}+ 2 \bar\Phi D\Phi D \bar{D}^2 \bar{\Phi} + \Phi\bar\Phi D^2\bar D^2 \bar\Phi\right) \delta\Phi, \\
\nonumber\\
 \int d^4\theta \,\, \delta \mathcal{O}_0 &=\int d^4\theta \,\, \delta(D \Phi D\Phi)\bar D \bar\Phi \bar D  \bar\Phi = -2 \int d^4\theta \,\, \left( \vphantom{\left(\bar{\Phi}\right)^2}  D^2 \Phi \bar{D}\bar{\Phi}\bar D \bar \Phi+ 2 D \Phi D \bar D \bar \Phi \bar{D} \bar{\Phi}\right) \delta\Phi \nonumber\\
&= \frac{1}{2} \int d^2\theta \,\, \bar D^2 \left( \vphantom{\left(\bar{\Phi}\right)^2}  D^2 \Phi \bar{D}\bar{\Phi}\bar D \bar \Phi+ 2 D \Phi D \bar D \bar \Phi \bar{D} \bar{\Phi}\right) \delta\Phi,
\end{align}
where the notation $\bar D \bar D$ indicates that, contrary to $\bar D ^2=\bar D_{\dot \alpha} \bar D^{\dot \alpha}$, the spinor indices are not contracted between the two covariant derivatives but with the other ones appearing in the same operator. 

The equation of motion for the generic theory of \cref{lagrangianSuperspace} thus reads:
\begin{equation}
-\frac{1}{4}\bar{D}^2 \bar{\Phi} + \frac{\partial W}{\partial \Phi}(\Phi)+ \epsilon^4 \left(\alpha \delta \tilde{\mathcal{O}}_0-\beta \delta \tilde{\mathcal{O}}_2 + \gamma (\delta \tilde{\mathcal{O}}_1+\delta \tilde{\mathcal{O}}_2\right)_{\theta\theta} = O(\epsilon^5). \label{fulleom4deriv}
\end{equation}

On a generic chiral superfield $\Psi$, we have according to \cref{stDvsSUSYD}:
\begin{equation}
\bar{D} D \Psi \sim \partial^{\mu} \Psi,
\end{equation}
such that the first step of the JLM procedure can be applied by replacing space-time derivatives by the $\bar D D$ operator on the zeroth order equations of motion \cref{0orderEOM}, namely by writing:
\begin{align}
&\epsilon^4 \left(-\frac{1}{4}\bar{D}^2 \bar{\Phi} + \frac{\partial W}{\partial \Phi}(\Phi)\right)=O(\epsilon^5),\nonumber  \\
&\epsilon^4 \left(-\frac{1}{4} \bar D D \bar{D}^2 \bar{\Phi} + \bar D D  \frac{\partial W}{\partial \Phi}(\Phi)\right)=O(\epsilon^5),\\
&\quad \vdots \nonumber 
\end{align}
The other constraints are then obtained following the procedure described earlier, on the full equation of motion \eqref{fulleom4deriv}. When only the first order is needed, the tree-level equation of motion is sufficient. Once all the HD operators are expressed in terms of $D^2\Phi$ or $\bar{D}^2\bar\Phi$, it is then possible to eliminate higher-order derivatives, making use of \cref{0orderEOM}. Not all HD operators can be written as covariant derivatives of these superfields, even after partial integration, for reasons similar to the ones explained in the non-supersymmetric case below \cref{e43}.

\subsubsection{Treating the SUSY HD Lagrangian at order $O(\epsilon^2)$}

For the higher-derivative Lagrangian \eqref{lagrangianSuperspace} at order $O(\epsilon^2)$, we can use directly the tree-level eom in the Lagrangian. For the $\mathcal{O}_1$ and $\mathcal{O}_2$ operators it gives:
\begin{align}
\epsilon^4  \int d^4\theta \mathcal{O}_2& = \epsilon^4  \int d^4\theta \Phi D^2 \Phi \bar D \bar \Phi \bar D \bar \Phi = 4 \epsilon^4  \int d^4  \theta \, \Phi \bar W'(\bar\Phi) \bar D \bar \Phi \bar D \bar \Phi = 4 \epsilon^4  \int d^4 \theta \,\Phi \bar D ( \bar W(\bar\Phi)) \bar D \bar \Phi \nonumber \\
&= - 4 \epsilon^4  \int d^4 \theta \,\Phi  \bar W(\bar\Phi) \bar D^2 \bar \Phi =  - 16 \epsilon^4  \int d^4 \theta \,\Phi W'(\Phi)  \bar W(\bar\Phi) + O(\epsilon^5), \\
\epsilon^4  \int d^4\theta \mathcal{O}_1& = \epsilon^4  \int d^4\theta \Phi \bar \Phi D^2 \Phi \bar D^2 \bar \Phi  = 16 \epsilon^4  \int d^4\theta \Phi \bar \Phi W'(\Phi)\bar{W}'(\bar\Phi) + O(\epsilon^5),
\end{align}
while the $\mathcal{O}_0$ operator does not contain terms proportional to the tree-level eom. From the two above expressions, it is easy to extract the component Lagrangian, by using the expression of the top component of the product of  chiral and anti-chiral superfields:
\begin{align}
\int d^4\theta  f(\Phi) \bar{g}(\bar \Phi)= f'(\phi) \bar{g}'(\bar \phi)\left\{F\bar F - \p_{\mu} \phi \p^{\mu} \bar\phi\right\} + {\rm fermions} + {\rm tot.} \, .\label{productchiralantichi}
\end{align}
We thus have that the following bosonic expansions:
\begin{align}
&  \frac{\epsilon^4}{16} \int d^4\theta {\mathcal{O}_2+\bar{\mathcal{O}}_2}=-\epsilon^4 (|W'|^2+\phi W''\bar W') \left\{F\bar F - \p_{\mu}\phi\p^{\mu}\bar\phi\right\}+ c.c. + O(\epsilon^5),\\
 &\frac{\epsilon^4}{8} \int d^4\theta {\mathcal{O}_1}
=2\epsilon^4 |W'(\phi) + \phi W''(\phi)|^2 \left\{F\bar F  -\p_{\mu}\phi\p^{\mu}\bar\phi\right\} + O(\epsilon^5) .
\end{align}
The operator $\mathcal{O}_0$ does not contain any higher-derivative terms, hence it does not need to be treated. We recall its bosonic component expansion: 
\begin{equation}
\frac{\epsilon^4}{16} \int d^4 \mathcal{O}_0=\epsilon^4 \int d^4\theta D\Phi D\Phi \bar D\bar\Phi \bar D\bar\Phi = \epsilon^4\left\{\p_{\mu}\phi\p^{\mu}\phi \p_{\nu}\bar \phi\p^{\nu}\bar \phi -2 F\bar F \p_{\mu}\phi\p^{\mu}\bar\phi + |F|^4\right\}\ .
\end{equation}
Putting everything together, we find that the higher-derivative Lagrangian $\mathcal{L}_{HD}$ of \eqref{lagrangianSuperspace} reads:
\begin{align}
\mathcal{L}_{HD}&=\epsilon^4 \left\{\alpha |F|^4 + \alpha (\p\phi)^2(\p\bar\phi)^2 \right.\nonumber\\
 &\left. - (2 \alpha F\bar F +2\beta |W'|^2 +(\beta+\gamma)(\phi W''\bar W'+\bar \phi \bar W'' W') + 2\gamma |\phi|^2|W''|^2)\p_{\mu}{\phi}\p^{\mu}\bar\phi\right.\nonumber\\
 &+\left. F\bar F   \left(2\beta |W'|^2+ (\beta+\gamma)(\phi W''\bar W'+\bar \phi \bar W'' W') + 2\gamma |\phi|^2|W''|^2\right)\right\} + O(\epsilon^5).
\end{align}
One can now integrate out the auxiliary field $F$ using its algebraic equation of motion. The tree level $F=-\bar W' + O(\epsilon^4)$ is once again sufficient, and using it in the above Lagrangian yields the Lagrangian \eqref{finalLcomponents}, previously derived fully in component form.

\subsection{Relation with the standard procedure}\label{comparisonwithadditionalfields}

We now shortly make a parallel between our treatment of higher-derivative operators in SUSY theories and other treatments made in the past \cite{Antoniadis:2007xc,Dudas:2015vka}. 

Already at the level of higher-derivative non-supersymmetric field theories, the standard procedure \cite{Ostrogradsky:1850fid,Langlois:2015cwa} to identify ghost degrees of freedom consists in introducing additional fields in order to write the theory in terms of an on-shell equivalent two-derivative one. This is exactly what was done at the end of the simple example of \cref{basicExampleHDtheory}. In the two-derivative multi-scalar theories, one can then identify the ghost modes by diagonalising the kinetic terms. It might not be possible in general to diagonalise the mass matrix simultaneously. In an EFT treatment where the higher-derivative terms are suppressed, the mass of the ghost states will be higher than those of the well-behaved ones. Integrating out the ghost modes then gives an effective two-derivative Lagrangian for the well-behaved modes.

In \cite{Dudas:2015vka}, this search for new degrees of freedom was done directly at the level of superfields, by introducing two new chiral superfields $\Psi_1$ and $\Psi_2$, for each chiral superfield appearing with a $\Box$ operator. Of these two new superfields, one comes from the superfield constraint $\Psi_1-\frac1m \bar D^2\bar\Phi=0$, and the other from the associated Lagrange multiplier: $\mathcal L\supset\int d^2\theta \,\Psi_2(\Psi_1-\frac1m \bar D^2\bar\Phi)+h.c.$. These new chiral superfields $\Psi_1$ and $\Psi_2$ account for the two new degrees of freedom generated by $\Box\phi$ and the dynamical part of $F$, both encountered in the field component expansion of the original $\Box\Phi$.

In the treatment we described in the previous sections, the ghost modes are not explicitly identified. Instead, the higher-derivative operators are directly eliminated from the low-energy theory by field redefinitions or by the use of equations of motion at a certain order of approximation. It avoids  the introduction of the new fields, diagonalisation of kinetic terms and integration of the heavy ghost modes, a task which can be tedious.

\section{Some applications of higher-derivative supersymmetric EFTs}\label{sec:Applications}

In this section, we apply the methods outlined above to physically interesting systems where HD terms play a prominent role: an extension of the MSSM, the SUSY embedding of the DBI action and the effective action of moduli in string compactifications. In the first case, we mostly review an example of treatment of HD operators in SUSY theories, whereas in the last two cases we are able to offer new insights (and corrections) on old results.

\subsection{MSSM with HD operators}

Higher-derivative operators of mass dimension five and six have been used to extend the MSSM and to correct the physical parameters \cite{Antoniadis:2008es,Antoniadis:2009rn}. We show here how such models naturally fit in the description we have given so far. We do not write the exact dimension-five and dimension-six higher-derivative operators used in the extension of the MSSM but rather study a simple scalar Lagrangian including the operators of interest. The main goal is to give a first application of the above methods in superspace and give a particular example where plugging the JLM constraints directly in the Lagrangian fails to reproduce the correct effective theory. This discussion closely follows the model studied in \cite{Dudas:2015vka} with Lagrangian:
\begin{align}
\mathcal{L}&=\int d^4\theta \left\{ \Phi\bar\Phi + \frac{\rho}{M^2} \bar\Phi\Box\Phi \right\}+ \left(\int d^2\theta \left \{ W(\Phi) + \frac{\sigma}{M}\Phi\Box\Phi \right\} + h.c.\right) + O(1/M^3) \label{MSSMHD} \\
&=\int d^4\theta \left\{ \Phi\bar\Phi - \frac{\rho}{16 M^2} \bar\Phi\bar D^2 D^2 \Phi \right\} + \left(\int d^2\theta \left\{ W(\Phi)- \frac{\sigma}{16 M}\Phi\bar D^2 D^2 \Phi\right\}  + h.c. \right) + O(1/M^3). \nonumber
\end{align}
Varying with respect to $\Phi$, using \cref{ibpsupefield} together with the chirality of $\Phi$ and integrating by parts, one obtains the following equation of motion:
\begin{equation}
-\frac14 \bar D^2 \bar \Phi + \frac{\partial W}{\partial \Phi} - \frac{\sigma}{8M}\bar D^2 D^2 \Phi + \frac{1}{4}\frac{\rho}{16 M^2} \bar D^2 D^2 \bar D^2 \bar \Phi =0. \label{eomSUSYDudas}
\end{equation}
We can then apply the reduction of order/JLM procedure defining $\epsilon=1/M$ and working at order $O(\epsilon^3)$. 

\subsubsection{$O(\epsilon^2)$ correction to the K\"ahler potential} We first study the case with $\sigma=0$. The HD operator reduces to an $O(\epsilon^2)$ correction to the K\"ahler potential. The Lagrangian indeed reads:
\begin{align}
\mathcal{L}&=\int d^4\theta \left\{ \Phi\bar\Phi - \epsilon^2 \frac{\rho}{16} \bar\Phi\bar D^2 D^2 \Phi \right\} + \left(\int d^2\theta \,\, W(\Phi)  + h.c. \right) + O(\epsilon^3). \label{Lcase1}
\end{align}
The first constraints of the JLM procedure, obtained at order $k=2$  by multiplying the equation of motion \eqref{eomSUSYDudas} by $\epsilon^2$, and differentiating,  thus read:
\begin{align}
&\epsilon^2\left(-\frac14 \bar D^2 \bar \Phi + \frac{\partial W}{\partial \Phi} \right) = O(\epsilon^3), \label{case1cons1} \\
&\epsilon^2 \left(-\frac14 D^2\bar D^2 \bar \Phi + D^2 \frac{\partial W}{\partial \Phi} \right) = O(\epsilon^3) \label{case1cons2}.
\end{align}
There are now two ways to proceed. Strictly following the JLM procedure, one should expand the second term of \cref{case1cons2} and replace terms of the form $D^2 \Phi$, including second derivatives of the scalar field, using the first constraint \eqref{case1cons1}. Another way is to directly plug the (hermitian conjugate of the) constraint \eqref{case1cons2} in the Lagrangian \eqref{Lcase1} and integrate by parts:
\begin{align}
\mathcal{L}&=\int d^4\theta \left\{ \Phi\bar\Phi - \epsilon^2 \frac{\rho}{4} \bar\Phi \bar D^2 \frac{\partial \bar W}{\partial \bar \Phi} \right\} + \left(\int d^2\theta \,\, W(\Phi)  + h.c. \right) + O(\epsilon^3) \nonumber\\
&=\int d^4\theta \left\{ \Phi\bar\Phi - \epsilon^2 \frac{\rho}{4} ( \bar D^2 \bar\Phi ) \frac{\partial \bar W}{\partial \bar \Phi} \right\} + \left(\int d^2\theta \,\, W(\Phi)  + h.c. \right) + O(\epsilon^3) \nonumber \\
&=\int d^4\theta \left\{ \Phi\bar\Phi - \epsilon^2 {\rho} \left| \frac{\partial \bar W}{\partial \bar \Phi} \right|^2  \right\} + \left(\int d^2\theta \,\, W(\Phi)  + h.c. \right) + O(\epsilon^3). 
\end{align}
This is exactly the Lagrangian obtained in \cite{Dudas:2015vka} by superspace  field redefinitions or via the standard procedure, namely by introducing additional fields accounting for the additional degrees of freedom and integrating them out.

\subsubsection{$O(\epsilon)$ correction to the  Superpotential} \label{MSSMcorrectionsuperpot}

We next consider the case $\rho=0$. This corresponds to an $O(\epsilon)$ correction to the superpotential. The Lagrangian thus reads:
\begin{equation}
\mathcal{L}=\int d^4\theta \Phi\bar\Phi + \left(\int d^2\theta \left\{ W(\Phi)- \epsilon \frac{\sigma}{16}\Phi\bar D^2 D^2 \Phi\right\}  + h.c. \right) + O(\epsilon^3). \label{Lcase2}
\end{equation}
This case is more involved than the previous one, as we expect that using the JLM constraints directly in the Lagrangian does not lead to the correct effective Lagrangian. According to the last paragraph of \cref{fieldredefs}, one can only do this to eliminate terms of order $O(\epsilon^k)$ when working with a Lagrangian approximated at $O(\epsilon^{n+1})$ with $2k\geq n+1$. This is not the case here as we consider  $O(\epsilon)$ corrections but keep working at order $O(\epsilon^3)$. This was already noticed in \cite{Antoniadis:2009rn,Dudas:2015vka}. 

We quickly show below the Lagrangian obtain from the wrong procedure, namely using directly the constraints in the Lagrangian. The first constraints are identical to those of the previous case, shown in \cref{case1cons1,case1cons2}, supplemented with a third one obtained by differentiating once more:
\begin{align}
&\epsilon^2 \left(-\frac14 \bar D^2 D^2\bar D^2 \bar \Phi + \bar D^2 D^2 \frac{\partial W}{\partial \Phi} \right) = O(\epsilon^3) \label{case1cons3}.
\end{align}
The secondary constraints are obtained by multiplying the equation of motion \eqref{eomSUSYDudas} by $\epsilon$ and differentiating. They read:
\begin{align}
&\epsilon \left( -\frac14 \bar D^2 \bar \Phi + \frac{\partial W}{\partial \Phi} - \epsilon \frac{\sigma}{8}\bar D^2 D^2 \Phi \right)=O(\epsilon^3), \nonumber\\
& \epsilon \left( -\frac14 D^2 \bar D^2 \bar \Phi + D^2 \frac{\partial W}{\partial \Phi} - \epsilon \frac{\sigma}{8} D^2 \bar D^2 D^2 \Phi  \right)=O(\epsilon^3).
\end{align}
Using the primary constraints \cref{case1cons1,case1cons2,case1cons3}, these last constraints shall be rewritten as:
\begin{align}
&\epsilon \left( -\frac14 \bar D^2 \bar \Phi + \frac{\partial W}{\partial \Phi} - \epsilon \frac{\sigma}{2}\bar D^2 \frac{\partial \bar W}{\partial \bar \Phi} \right)=O(\epsilon^3), \nonumber\\
&\epsilon \left( -\frac14 D^2 \bar D^2 \bar \Phi + D^2 \frac{\partial W}{\partial \Phi} - \epsilon \frac{\sigma}{2} D^2 \bar D^2 \frac{\partial \bar W}{\partial \bar \Phi} \right)=O(\epsilon^3).
\end{align}
Plugging the last constraint in the Lagrangian \eqref{Lcase2} and using \cref{ibpsupefield} leads to:
\begin{align}
\mathcal{L}_{wrong}
&=\int d^4\theta \Phi\bar\Phi + \left( \int d^2\theta W(\Phi) +\int d^4\theta \epsilon \sigma \Phi \frac{\partial \bar W}{\partial \bar \Phi} - \int d^4\theta \epsilon^2 \frac{\sigma^2}{2}\Phi D^2 \frac{\partial W}{\partial \Phi} +h.c.\right) +O(\epsilon^3). \label{case2firstform}
\end{align}
This third and last term of \cref{case2firstform} can be rewritten using superspace integration by parts and the primary constraint \eqref{case1cons1} and lead to the equivalent Lagrangian:
\begin{equation}
\mathcal{L}_{wrong}=\int d^4\theta \left(\Phi\bar\Phi - 8 \epsilon^2 \sigma^2 \left|\frac{\partial W}{\partial \Phi}\right|^2 \right)+\left(\int d^2\theta \left\{ W(\Phi) - \epsilon \sigma {W'}^2\right\} + h.c.\right) +O(\epsilon^3). \label{Lwrong}
\end{equation}
This result does not agree with the one obtained using field redefinitions in superspace or by including additional fields and integrating them out. Indeed, the correct result is \cite{Dudas:2015vka}:
\begin{equation}
\mathcal{L}=\int d^4\theta \left(\Phi\bar\Phi - 4 \epsilon^2 \sigma^2 \left|\frac{\partial W}{\partial \Phi}\right|^2 \right)+\left(\int d^2\theta \left\{ W(\Phi) - \epsilon \sigma  {W'}^2 + 2 \epsilon^2 \sigma^2 {W'}^2 W'' \right\} + h.c.\right) \label{LagrangianGD2} +O(\epsilon^3).
\end{equation}
This shows explicitly that one should use the prescription of the last paragraph of \cref{fieldredefs} to know when the equivalent effective theory can be obtained  using the constraints directly in the Lagrangian. We recall that when this is not the case, one should use the constraints recursively in order to obtain the physical equation of motion, as explained in detail in \cref{JLMconstraints}. Solutions of this effective eom are identical to the ones derived from the equivalent two-derivative Lagrangian obtained using field redefinitions. 

In \cref{HDMSSMcomponents} we show, working in components, that the difference between the two above Lagrangians is exactly made of the second order terms of the field redefinitions. These terms are neglected when plugging the JLM constraints in the Lagrangian. 

\subsection{Supersymmetric embedding of the chiral DBI action}\label{sectionDBI}
In this section, armed with the techniques developed previously to analyse higher-derivative SUSY EFTs, we revisit the SUSY description of the DBI action proposed in \cite{Bielleman:2016olv}.

The bosonic part of the DBI action describing a D$p$ brane in a type IIB toroidal orbifold compactification can be expressed as (see e.g. \cite{Bielleman:2016olv}): 
\begin{align}
\mathcal{L}_{DBI}=-\frac{\mu_p}{g_s} \, V_{p-3} f(\phi_i) \sqrt{-\det\{g_{\mu\nu}+Z \sigma^2 (\partial_{\mu}\phi_i\partial_{\nu} \bar{\phi}_i+\partial_{\mu}\bar{\phi}_i\partial_{\nu} \phi_i)\}},
\end{align}
where $\sigma=2\pi\alpha'$, $V_{p-3}$ is the volume of the cycle wrapped by the brane, $\mu_p=1/(2\pi)^p\sqrt{\alpha'^{p+1}}$ and $Z$ is a possible wrapping factor. The scalars $\phi_i$ correspond to the moduli parameterising the transverse position of the brane. The determinant inside the square root can be expanded and reads exactly:
\begin{align}
\Delta(\phi_i)&\equiv - \det\{g_{\mu\nu}+Z \sigma^2 (\partial_{\mu}\phi_i\partial_{\nu} \bar{\phi}_i+\partial_{\mu}\phi_i\partial_{\nu} \bar{\phi}_i)\}\nonumber\\
&=1+2 Z\sigma^2 \partial_{\mu}\phi_i\partial^{\mu}\bar{\phi}_i +Z^2\sigma^4 \left( 2(\partial_{\mu}\phi_i\partial^{\mu}\bar{\phi}_i)^2 - \p_{\mu}\phi_i \p^{\mu}\phi_j \p_{\nu}\bar \phi_i \p^{\nu}\bar \phi_j -\p_{\mu}\phi_i \p^{\mu}\bar \phi_j \p_{\nu}\phi_j \p^{\nu}\bar \phi_i\right),
\end{align}
so that the DBI Lagrangian reads:
\begin{equation}
\mathcal{L}_{DBI}=-\frac{\mu_p}{g_s} \, V_{p-3} f(\phi_i)\sqrt{\Delta(\phi_i)}.
\end{equation}
To obtain an approximate Minkowski vacuum, we consider the presence of an orientifold plane, described by the Lagrangian:
\begin{equation}
\mathcal{L}_{Op}=\frac{\mu_p}{g_s} \, V_{p-3}.
\end{equation}
Finally, the brane Lagrangian also includes the Chern-Simons term, which in a SUSY setup reads \cite{Bielleman:2016olv}:
\begin{equation}
\mathcal{L}_{CS}=-\frac{\mu_p}{g_s}\, V_{p-3} (f(\phi_i)-1),
\end{equation}
so that the total 4d Lagrangian is:
\begin{equation}
\mathcal{L}=\frac{\mu_p}{g_s}\,V_{p-3} \left[(2-f(\phi_i))-f(\phi_i)\sqrt{\Delta(\phi_i)}\right].
\end{equation}
We now consider the case with only one complex scalar modulus $\phi$. It corresponds to a two-dimensional transverse space, hence to a $D7$ brane. In that case, the determinant $\Delta(\phi)$ reads:
\begin{equation}
\Delta(\phi)=1+2 Z\sigma^2 \partial_{\mu}\phi\partial^{\mu}\bar{\phi} +Z^2\sigma^4 \left( (\partial_{\mu}\phi\partial^{\mu}\bar{\phi})^2 - \p_{\mu}\phi \p^{\mu}\phi \p_{\nu}\bar \phi \p^{\nu}\bar \phi\right).
\end{equation}
Using $\mu_7 = 1/(8 \pi^3 \sigma^4)$ and defining the dimensionless volume $\mathcal V_4 = V_4/\sigma^2$, we can write the Lagrangian as:
\begin{align}
\mathcal{L}&=\frac{\mathcal{V}_{4}}{(2\pi)^3 g_s}\frac{1}{\sigma^2} \left[2-f(\phi)-f(\phi)\sqrt{1+2 Z\sigma^2 \partial_{\mu}\phi\partial^{\mu}\bar{\phi} +Z^2\sigma^4 \left( (\partial_{\mu}\phi\partial^{\mu}\bar{\phi})^2 - (\p\phi)^2(\p\bar\phi)^2\right)}\right].
\end{align}
Dropping the warp factor $Z$ and expanding the square root allows to rewrite the action as
\begin{align}
\mathcal{L}&=\frac{\mathcal{V}_{4}}{(2\pi)^3 g_s \sigma^2} \, \left[2(1-f(\phi))-f(\phi) \sigma^2 \p_{\mu}\phi\p^{\mu}\bar\phi+ \frac 12 f(\phi)\sigma^4  (\p\phi)^2(\p\bar\phi)^2 + \ldots \right] \nonumber\\
&= -V(\phi) - a \left(1+\frac{\sigma^2}{2 a} V(\phi)\right)\partial_{\mu}\phi\partial^{\mu}\bar{\phi}+ \frac{a}{2} \sigma^2 (\p\phi)^2(\p\bar\phi)^2 + O(\sigma^4),
\end{align}
with:
\begin{equation}
V(\phi)=\frac{2a }{\sigma^2} (f(\phi)-1), \quad a\equiv \frac{\mathcal{V}_{4}}{(2\pi)^3 g_s}.
\end{equation}

 Rescaling $\tilde \phi=\sqrt{a}\phi$ and defining the new expansion parameter $ \epsilon^4\equiv {\sigma^2}/{a}$, we obtain the Lagrangian:
 \begin{equation}
 \mathcal{L}=-\tilde{V}-\left(1+\frac{\epsilon^4}{2} \tilde{V}\right)|\partial \tilde{\phi}|^{2}+\frac{1}{2}\epsilon^4(\partial \tilde{\phi})^{2}(\partial \bar{\tilde{\phi}})^{2}+O(\epsilon^{8}), \label{LDBIexp}\, \quad \tilde{V}(\tilde\phi)\equiv V\Big({\tilde\phi}/{\sqrt{a}}\Big).
 \end{equation}
Note that the scalar potential does not receive any correction proportional to $\epsilon$. This feature plays a pivotal role in the search for a SUSY description of this action, as shown hereafter.

We indeed look for a SUSY description of the physics of the above brane action by comparing with the Lagrangian \eqref{finalLcomponents} and adjusting the three free coefficients $\alpha$,$\beta$ and $\gamma$. It amounts to identify the linear combination of higher-derivative operators leading to the brane action. In particular, we look for parameters eliminating the correction to the scalar potential in \cref{finalLcomponents,correctedVtot}. The functions introduced there should thus satisfy:
\begin{equation}
    f(\phi,\bar \phi)=\alpha V(\phi,\bar \phi),\label{PDE1}
\end{equation}
leading to a Lagrangian of the form:
\begin{equation}
\mathcal{L}=-V-(1+\epsilon^{4}\alpha V)|\partial \phi|^{2}+\epsilon^{4} \alpha(\partial \phi)^{2}(\partial \bar \phi)^{2}\,\,.
\end{equation}
Notice how the exact matching of the coefficients in \eqref{LDBIexp} is recovered by setting $\alpha=1/2$ for a potential subject to \cref{PDE1}.

It is interesting to investigate what constraints are placed on the scalar potential by requiring that the HD action matches the  form of the brane action. These constraints arise from the correct treatment of terms involving spacetime derivatives of the $F$ auxiliary field and  were absent in \cite{Bielleman:2016grv,Bielleman:2016olv}, where these were set to zero. The potential $V$ must solve the partial differential equation \eqref{PDE1}, that explicitly reads:
\begin{equation}
\left(\alpha+2\beta\right)V+2\gamma|\phi|^{2}V_{\phi\bar \phi}+\left(\beta+\gamma\right)\left(\phi V_{\phi}+\bar \phi V_{\bar \phi}\right)=0\label{PDE2}\,\,.
\end{equation}
This equation is solved by:
\begin{equation}
    V(\phi,\bar \phi)=2^{\frac{n}{2}}v_0 |\phi|^n,
\end{equation}
with the power $n$ depending on the coefficients $\alpha$, $\beta$, $\gamma$ of the HD operators through
\begin{equation}
    \gamma n^2 + 2(\beta+\gamma)n + 2(\alpha+2\beta)=0\,\,.
\end{equation}
This constraint admits both positive and negative integer solutions for  $n$, depending on the underlying parameters $\alpha,\beta$ and $\gamma$. Such scalar potential can be obtained in global SUSY from the following superpotential:  
\begin{equation}
	 W(\Phi)=\dfrac{2^{\frac{n}{4}+1}\mathit{v}_{0}^{\frac{1}{2}}}{n+2}\,\Phi^{\frac{n}{2}+1}\label{e3.113}\,\,.
\end{equation}
To conclude, the explicit form for the supersymmetric Lagrangian that allows to recover the characteristic features of the brane Lagrangian at its first-order expansion is given by the superpotential \eqref{e3.113} together with the corrected K\"ahler potential:
\begin{align}
K=|\Phi|^{2} +\frac{\epsilon^4}{16}\left[{\alpha} D\Phi D\Phi\bar{D}\bar{\Phi}\bar{D}\bar{\Phi}+2{\gamma}|\Phi|^{2}D^{2}\Phi\bar{D}^{2}\bar{\Phi}+(\beta-\gamma)\left(\Phi D^2\Phi(\bar{D}\bar{\Phi})^2+h.c.\right)\right] +O(\epsilon^{8}).
\end{align}
We stress that this result differs from that of \cite{Bielleman:2016grv,Bielleman:2016olv} where the terms containing derivatives of $F$ were set to zero by hand.

\subsection{Moduli stabilization in 4d $\mathcal{N}=1$ SUSY String reductions}

In this section we reevaluate the HD corrections to the scalar potential for moduli fields in IIB string compactifications \cite{Ciupke:2015msa} originating from the dimensional reduction of $\alpha'^3 R^4$ terms of the 10d action. Such terms, besides the well-known correction to the K\"ahler potential  originating from a correction to the scalar kinetic term \cite{Antoniadis:1997eg,Becker:2002nn}, also give rise to K\"ahler moduli dependent four-derivative interactions which, through SUSY, generate extra terms in the scalar potential. 

The specific coefficients for the $|F|^4$ terms and the additional four-derivative operators were computed in \cite{Grimm:2017okk}.
We consider the case of one K\"ahler modulus $u=\log(\mathcal{V}-\mathcal{V}_0)$, where $\mathcal{V}$ denotes the internal volume and $\mathcal{V}_0$ its vacuum expectation value. The second-derivative terms then take the standard form \footnote{We omit the  $\alpha'^3$ correction to the kinetic term \cite{Antoniadis:1997eg,Becker:2002nn}  as it does not play a role in this discussion.}:
\begin{equation}
S_{IIB}^{(2)}=\frac{\mathcal{V}_0 M_s^2}{2
}\int d^4 x \sqrt{-g}\left( R -\frac{2}{3} \partial_{\mu}u \partial^{\mu} u \right),
\end{equation}
while up to partial integration, the bosonic four-derivative terms read:
\begin{equation}
S_{IIB}^{(4)}= \beta \int d^4x \sqrt{-g} \left\{ a\ (\partial_{\mu}u)^4 + b\ (\Box u)^2 + c\ \Box u \partial_{\mu} u \partial^{\mu} u\right\} \, ,\label{alpha3hd}
\end{equation}
where $\beta$ is a dimensionless parameter dependent on the compact space's geometry and $a,b$ and $c$ are rational numbers. The authors of \cite{Ciupke:2015msa} focus only on the first term of \cref{alpha3hd}, arguing that it suffices to get the functional dependence and that the remaining ones would give rise to ghosts. Furthermore, when looking for the correct SUSY operator to describe this action they consider exclusively the $\mathcal{O}_0$ operator, introduced in eq. \eqref{4fields4deriv}, arguing that it is "the unique ghost-free operator", with all the others giving rise to dynamical $F$ fields. Given the perturbative treatment developed in this paper we know that not to be the case.  A more systematic approach was employed in \cite{Grimm:2017okk}, where these additional terms of \cref{alpha3hd} were eliminated via a field redefinition in the kinetic part of the Lagrangian. However, when the action is supplemented by a scalar potential the field redefinitions performed exclusively in the kinetic part of the action miss some of the corrections to the scalar potential as we have explicitly shown in the global SUSY result of \cref{modulistabcorrectedscalarpot}. We are therefore compelled to revisit this in light of the methods outlined before.
Using that $1/\kappa_{10}^2 \sim {\alpha'}^4 \sim M_P^2/\mathcal{V}_0$, and $\alpha'=1/M_s^2$, one can rewrite the above action in terms of the canonically normalized scalar $\varphi= \sqrt{\frac{2}{3}} u  M_P $ as:
\begin{align}
S_{IIB}=& \int d^4x \sqrt{-g} \left(\frac{M_P^2}{2} R -\frac{1}{2} \partial_{\mu}\varphi \partial^{\mu} \varphi \right) \nonumber\\
&+ \beta \int d^4x \sqrt{-g} \left\{ \frac{\tilde{a} }{M_P^4} (\partial_{\mu}\varphi)^4 + \frac{\tilde{b}}{M_P^2} (\Box \varphi)^2 +  \frac{\tilde{c}}{M_P^3}\Box \varphi \partial_{\mu} \varphi \partial^{\mu} \varphi\right\},
\label{alpha3hdphi}
\end{align}
where $\tilde{a}=9/4\ a$, $\tilde{b}=3/2\ b$ and $\tilde{c}=(3/2)^{3/2}\ c$. We see that even though all HD terms originate from the same ten dimensional $\alpha'^3$ correction, upon dimensional reduction and canonical normalisation they are suppressed by different powers of $M_P$. Clearly in an expansion in $1/M_P$ the leading term is $(\Box \varphi)^2$ rather than $(\partial \varphi)^4$ as considered both in \cite{Ciupke:2015msa} and \cite{Grimm:2017okk}.

We could introduce the following redefinition of the scalar $\varphi$, similar to the one used in \cite{Grimm:2017okk} for the dimensionless field $u$ :
\begin{equation}
\varphi \rightarrow \varphi - \beta \Big(\frac{\tilde b}{M_P^2} \Box \varphi + \frac{\tilde c}{M_P^3} \p_{\mu}\varphi \p^{\mu} \varphi \Big).
\end{equation}
It brings the above HD action to:
\begin{align}
S_{IIB}= \int d^4x \sqrt{-g} \left(\frac{M_P^2}{2} R -\frac{1}{2} \partial_{\mu}\varphi \partial^{\mu} \varphi \right) + \beta \int d^4x \sqrt{-g} \frac{\tilde{a} }{M_P^4} (\partial_{\mu}\varphi)^4 + O(\beta^2),
\end{align}
so that we see that it is indeed sufficient to keep only the $\tilde a$ term when keeping the leading term in the $\beta$ expansion. However, how should be clear from the rest of the paper, this is not true when a scalar potential is present, as is the case in a global SUSY embedding with non-vanishing superpotential. We thus refrain from using this field redefinition and study the action \eqref{alpha3hdphi}.

The four derivative operators correcting the K\"ahler potential with one chiral and one anti-chiral superfield were given in \cref{2fields4deriv}. Those with two chiral and one anti-chiral superfields were given in \cref{3fields4deriv1,3fields4deriv2}, and the ones with four superfields in \cref{4fields4deriv}.  A correction of the form:
\begin{align}
\delta \mathcal{K}&=\frac{\beta}{16} \left( \frac{\tilde{a}}{M_P^4}  \mathcal{O}_0 +\frac{\tilde{b}}{M_P^2}  \mathcal{N}_1 +\frac{\tilde{c}}{2M_P^3}  \mathcal{M}_2 + \frac{\tilde{c}}{2M_P^3} \bar{\mathcal{M}}_2  \right)\nonumber\\
&=\frac{\beta}{16}\left(\frac{\tilde{a}}{M_P^4}  D \Phi D \Phi\bar D\bar \Phi \bar D \bar \Phi + \frac{\tilde{b}}{M_P^2}  \bar \Phi \bar D^2 D^2 \Phi + \frac{\tilde{c}}{2M_P^3}  D\Phi D\Phi \bar D^2 \bar\Phi + \frac{\tilde{c}}{2M_P^3}  \bar D\bar \Phi \bar D\bar \Phi  D^2 \Phi \right) ,
\end{align}
would thus lead to:
\begin{align}
\mathcal{L}= & \, \phi\Box\bar\phi + F\bar F + F W' + \bar F \bar{W}' \nonumber\\
&+ \beta \left( \frac{\tilde{a}}{M_P^4}  (\partial \phi)^2(\partial \bar\phi)^2+  \frac{\tilde{b}}{M_P^2} \Box\phi \Box \bar \phi + \frac{\tilde{c}}{2M_P^3}(\partial_{\mu}\phi\p^{\mu} \phi\Box \bar \phi + \partial_{\mu}\bar \phi\p^{\mu} \bar \phi \Box  \phi  )\right. \nonumber\\
&\hspace{1.2cm} + \left. \frac{\tilde{a}}{M_P^4}\left(- 2 F\bar F \partial_{\mu}\phi\p^{\mu}\bar\phi + F^2 \bar F^2 \right) + \frac{\tilde{b}}{M_P^2}  F\Box \bar F+ \frac{\tilde{c}}{M_P^3} \left(\bar F \p_{\mu} \phi \p^{\mu} F + F \p_{\mu}\bar \phi \p^{\mu} \bar F\right) \right) \nonumber\\
& + O(\beta^2). \label{alpha3hdSUSY}
\end{align}
We see that, upon identifying $\varphi=Re(\phi)$, this reproduces the form of the HD terms of \cref{alpha3hdphi} and contains derivative terms for the auxiliary field, as expected. As described earlier, the Lagrangian can be treated as an EFT in $\beta$ and, either through field redefinitions or through the use of the perturbative constraints, it can be brought to the form
\begin{align}
\mathcal{L}&=\phi\Box\bar\phi + F\bar F + FW' +\bar F\bar W' + \beta \frac{\tilde{a}}{M_P^4} (\p \phi)^2(\p \bar \phi)^2 \nonumber\\
&\hspace{0.5cm} - \beta \left( \frac{2\tilde{a}}{M_P^4} F\bar F + \frac{\tilde{b}}{M_P^2} |W''|^2 - \frac{\tilde{c}}{2 M_P^3} (W'\bar W''+\bar W'  W'')\right)\p_{\mu}\phi\p^{\mu}\bar\phi \nonumber\\
&\hspace{0.5cm} + \beta \left( \frac{\tilde{a}} {M_P^4} F\bar F + \frac{\tilde{b}}{M_P^2}|W''|^2 - \frac{\tilde{c}}{2 M_P^3} (W'\bar W''+\bar W'  W'')\right) F\bar F + O(\beta^2)\ ,
\end{align}
where the auxiliary field $F$ appears algebraically and the only remaining HD term for $\phi$ is of the form $(\partial \phi)^2 (\partial \bar{\phi})^2$. After elimination of $F$ using $F=-\bar W' + O(\beta)$, the Lagrangian reads:
\begin{align}
\mathcal{L}&=\phi\Box\bar\phi -|W'|^2 +  \beta \frac{\tilde{a}}{M_P^4} (\p \phi)^2(\p \bar \phi)^2\nonumber\\
&\hspace{0.5cm}  - \beta \left( \frac{2\tilde{a}}{M_P^4}  |W'|^2 + \frac{\tilde{b}}{M_P^2}  |W''|^2 -  \frac{\tilde{c}}{2 M_P^3} ( W'\bar W''+\bar W'  W'')\right)\p_{\mu}\phi\p^{\mu}\bar\phi \nonumber\\
&\hspace{0.5cm} + \beta |W'|^2  \left( \vphantom{1^{1^1}} \frac{\tilde{a}}{M_P^4}  |W'|^2 + \frac{\tilde{b}}{M_P^2}  |W''|^2 - \frac{\tilde{c}}{2 M_P^3}(  W'\bar W''+\bar W'  W'')\right) + O(\beta^2), \label{modulistabcorrectedscalarpot}
\end{align}
so that the four-derivative correction to the scalar potential is of the form:
\begin{equation}
V_{HD}= \beta |W'|^2  \left( \vphantom{1^{1^1}} \frac{\tilde{a}}{M_P^4}  |W'|^2 + \frac{\tilde{b}}{M_P^2}  |W''|^2 - \frac{\tilde{c}}{2 M_P^3}(  W'\bar W''+\bar W'  W'')\right)\ .
\end{equation}
Had we performed the field redefinitions to eliminate the $\tilde{b}$ and $\tilde{c}$ terms \emph{before} doing the SUSY embedding as in \cite{Grimm:2017okk} we would have obtained instead (by setting $\tilde{b}=\tilde{c}=0$ in \cref{alpha3hdSUSY})
\begin{align}
\mathcal{L}&= \phi\Box\bar\phi -  |W'|^2 \nonumber\\
&\hspace{0.5cm}+ \beta \frac{\tilde{a}}{M_P^4} \left( (\partial \phi)^2(\partial \bar\phi)^2- 2 |W'|^2  \partial_{\mu}\phi\p^{\mu}\bar\phi + |W'|^4 \right)\ ,
\end{align}
where, in accordance with \cite{Ciupke:2015msa}, one finds an $|F|^4=|W'|^4$ correction to the scalar potential on top of a correction to the kinetic term and a HD term. The authors showed that this term is embedded in SUGRA with multiple scalars $\Phi_i$ as:
\begin{equation}
V_{|F|^4}= e^{2K} T^{ij\bar k\bar l}D_i W D_j \bar D_{\bar k}\bar W \bar D_{\bar l} \bar W, \label{sugraF4}
\end{equation}
where $D_i=\p_i + K_i$ denote here K\"ahler covariant  derivatives. Here $T^{ij\bar k\bar l}$ is the multi-scalar field extension of the parameter $a$, which is thus related to the effective action \cref{alpha3hdSUSY}. In SUGRA the K\"ahler covariant derivatives lead to a correction to the scalar potential, even in presence of a constant superpotential $W=W_0$, of the form \cite{Ciupke:2015msa}:
\begin{equation}
V_{|F|^4}  =- \hat{\lambda} \frac{\Pi_i t^i}{\mathcal{V}^4} {|W_0|^4} \ .\label{F4corrsugra}
\end{equation}
We note from \cref{modulistabcorrectedscalarpot} that, in global SUSY and for a generic superpotential, the functional form of the correction to $V$ is different from those considered in \cite{Ciupke:2015msa}, with the $\tilde{b}$ term dominating the $1/M_P$ expansion. It is natural to ask whether these differences persist in the SUGRA  embedding. If this would be the case there could be interesting consequences for the K\"ahler moduli vacuum structure where subleading corrections often play a crucial role in moduli stabilisation due to the no-scale structure the leading order F-term scalar potential is either flat or of run-away form. We will return to this issue in the future.



\section{Discussion}

In this work we have studied higher-derivative perturbations of supersymmetric EFTs, focusing on the methods to systematically eliminate ghosts from the spectrum. We have done so by applying two equivalent methods: JLM/reduction of order in the eom and field redefinitions in the action. 

This investigation was prompted by the desire to clarify the status of the auxiliary fields, $F$, in the EFT in the presence of higher-derivative terms. HD actions generically give rise to higher-order eom for scalar components that, in non-supersymmetric theories can be eliminated by both of the above methods. In the presence of SUSY, the higher-derivative terms will in general also induce non-algebraic (i.e. differential) and non-linear eom for the auxiliary fields. A study of the literature reveals diverging approaches on how to deal with these ``dynamical" auxiliary fields: from the {\it ab initio} exclusion from the EFT of the HD terms that induce them \cite{Khoury:2010gb,Koehn:2012ar,Koehn:2012np,Ciupke:2015msa} to the {\it ad hoc} imposition of the constraint $\partial F=0$ based on EFT intuition \cite{Bielleman:2016grv,Bielleman:2016olv}. 

Through a systematic application of JLM/reduction of order or field redefinitions we have indeed shown that the auxiliary fields remain non-dynamical in the EFT. The elimination of terms containing its derivatives yields perturbative corrections to the action. This is unambiguous when the action is expanded in component fields but can also be worked out using superfields. We have indeed proposed a reduction of order procedure directly in $\mathcal{N}=1$ superspace.

We have applied these methods to the study of several phenomena, reviewing at first possible higher-derivative MSSM extensions \cite{Antoniadis:2008es,Antoniadis:2009rn} and then investigating the scalar sector of the SDBI action and the effective action of 4d $\mathcal{N}=1$ compactifications of type IIB string theory. 

In the first of these applications, we have shown that a particular combinations of four-derivative operators reproduce the first terms of the bosonic expansion of D7-brane actions. One peculiarity of the brane action is that the HD operators do not correct the scalar potential, a feature that constrains the SUSY embedding of said potential to be of monomial form. We produced a supersymmetric EFT with all these properties, namely an uncorrected scalar potential and the correct matching of the coefficient of the first higher-derivative term. It is described by a power-law superpotential and a corrected K\"ahler potential.

In our study of 4d $\mathcal{N}=1$ compactifications of type IIB string theory, we have argued that one should extend the  previous study \cite{Ciupke:2015msa} by including all the HD operators deriving from the reduction of the 10d effective action \cite{Grimm:2017okk}, not only the \emph{trivially} ghost-free one. Working in global SUSY, we showed that the elimination of the additional ``pathological" HD operators from the EFT gives rise to a ghost-free theory with a corrected scalar potential. We found that the corrections to the scalar potential coming from the previously neglected HD terms actually dominate in an $1/M_P$ expansion over the one considered in \cite{Ciupke:2015msa,Grimm:2017okk}. The form of these corrections in the global SUSY case can give intuition on their lift to SUGRA. One can crudely estimate the K\"ahler moduli dependence of the novel corrections by replacing derivatives with the K\"ahler covariant derivatives of supergravity. For a constant superpotential,  this dependence is then identical to the one of the usual $|F|^4$ corrections from the  \emph{trivially}  ghost-free operator. Should this ansatz be confirmed by a full SUGRA analysis, we would then conclude that including all HD operators descending from the 10d effective action merely modifies the $\lambda$ coefficient of the $|F|^4$ higher-derivative corrections, while keeping its functional form unchanged. If on the other hand the dependence in the K\"ahler moduli does not follow this naive guess,  the functional shape of the scalar potential of the EFT could be modified. Both possibilities can have important consequences for moduli stabilisation in models where HD corrections plays an important role, see eg \cite{Cicoli:2016chb,Cicoli:2023njy}. It is thus necessary to perform this analysis directly at the level of the supergravity theory and derive the exact form of the corrections to the EFT scalar potential  when including all the HD operators, for a fixed internal manifold, a task that we leave for future work.

\section*{Acknowledgements}

LP is supported by the fellowship LCF/BQ/DI22/11940039 from ``La Caixa" Foundation (ID 100010434).
FGP and OL are partly supported by the COST (European Cooperation in Science and Technology) Action COSMIC WISPers CA21106.

\appendix

\section{Quantization of higher-derivative Lagrangians}\label{AppendixQuantization}

In this appendix, we shortly review quantisation of higher-derivative Lagrangians, and refer to \cite{Weldon:2003by} for a more comprehensive review.  The canonical quantisation of the theory uses the Hamiltonian formalism, described by first splitting time and space derivatives in the Lagrangian:
\begin{equation}
\mathcal{L}=\frac{1}{2}\dot{\phi}^{2}-\frac{1}{2}(\nabla\phi)^{2}-\frac{1}{2}m^{2}\phi^{2}-\frac{1}{2}\epsilon^2 \ddot{\phi}^{2}-\frac{1}{2}\epsilon^2(\nabla^{2}\phi)^{2}+\epsilon^2 \ddot{\phi}\nabla^{2}\phi\,, \label{LagrangianHD1}
\end{equation}
and then defining the conjugate momenta. Lagrangian theories with higher-derivatives contain several conjugate momenta. We refer to \cite{Woodard:2015zca} for a review of Ostrogradsky's original theorem \cite{Ostrogradsky:1850fid} and methods. The two canonical momenta for the Lagrangian \eqref{LagrangianHD1} read:
\begin{align}
\Pi_{\phi}&=\sum_{k=0}^{1}(-1)^{k}\left( \frac{\partial}{\partial t}\right)^{k}\frac{\partial\mathcal{L}}{\partial \phi^{(k+1)}}=\frac{\partial\mathcal{L}}{\partial \dot{\phi}}-\frac{\partial}{\partial t}\frac{\partial\mathcal{L}}{\partial \ddot{\phi}}=\dot{\phi}+\epsilon^{2}\dddot{\phi}-\epsilon^{2}\nabla^{2}\dot{\phi}, \label{momentum1}\\
\Pi_{\dot{\phi}}&=\frac{\partial\mathcal{L}}{\partial \ddot{\phi}}=\epsilon^{2}\nabla^{2}\phi-\epsilon^{2}\ddot{\phi}. \label{momentum2}
\end{align}
The Hamiltonian density defined using these momenta is thus given by
\begin{align}
\mathcal{H}&=\Pi_{\phi}\dot{\phi}+\Pi_{\dot{\phi}}\ddot{\phi}(\phi,\dot{\phi},\Pi_{\dot{\phi}})-\mathcal{L}(\phi,\dot{\phi},\Pi_{\dot{\phi}}) \nonumber \\
&=\Pi_{\phi}\dot{\phi}+\Pi_{\dot{\phi}}\nabla^{2}\phi-\frac{1}{2\epsilon^2} \Pi_{\dot{\phi}}^{2}-\frac{1}{2}\dot{\phi}^{2}+\frac{1}{2}(\nabla\phi)^{2}+\frac{1}{2}m^{2}\phi^{2}\,\,. \label{HamiltonianHD1}
\end{align}
Note that in this case only $\ddot{\phi}$ is eliminated from the Hamiltonian, since $\dot{\phi}$ has become part of the configuration space. In higher-derivative theories, only the higher time-derivatives are not part of the configuration space and must be substituted in terms of the momenta.

After quantisation, the scalar field $\hat{\phi}(x)$ described by the Lagrangian \eqref{LagrangianHD1} can be expanded as:
\begin{align}
\hat{\phi}(\vec{x},t)=\int \frac{d^{3}p}{(2\pi)^{3}}\biggl[&\frac{1}{\sqrt{2\omega_{c\vec{p}}}}\left(\hat{c}_{\vec{p}}\,e^{-i(\omega_{c\vec{p}}t-\vec{p}\cdot\vec{x})}+\hat{c}^{\dagger}_{\vec{p}}\,e^{+i(\omega_{c\vec{p}}t-\vec{p}\cdot\vec{x})}\right)+\\\nonumber
&\frac{1}{\sqrt{2\omega_{d\vec{p}}}}\left(\hat{d}_{\vec{p}}\,e^{-i(\omega_{d\vec{p}}t-\vec{p}\cdot\vec{x})}+\hat{d}^{\dagger}_{\vec{p}}\,e^{+i(\omega_{d\vec{p}}t-\vec{p}\cdot\vec{x})}\right)\biggr],
\end{align}
where $\hat{c}_{\vec{p}}$, $\hat{c}^{\dagger}_{\vec{p}}$ (resp.  $\hat{d}_{\vec{p}}$, $\hat{d}^{\dagger}_{\vec{p}}$) are the annihilation and creation operators of a convergent (resp. divergent) mode of momentum $\vec{p}$ and energy $\omega_{c\vec{p}}$ (resp. $\omega_{d\vec{p}}$).

As in the canonical quantisation procedure, one imposes the commutation relations between the fields and their conjugate momenta
\begin{equation} [\phi(\vec{x},t),\Pi_{\phi}(\vec{y},t)]=i \delta^{(3)}(\vec{x}-\vec{y}),\qquad  [\dot{\phi}(\vec{x},t),\Pi_{\dot{\phi}}(\vec{y},t)]=i \delta^{(3)}(\vec{x}-\vec{y}). \end{equation}
This leads to the following commutation relations for the creation and annihilation operators:
\begin{align}
\left[\hat{c}_{\vec{p}},\hat{c}^{\dagger}_{\vec{q}}\right]&=(2\pi)^{3}\delta^{(3)}(\vec{p}-\vec{q}),\label{com1}\\
\left[\hat{d}_{\vec{p}},\hat{d}^{\dagger}_{\vec{q}}\right]&=(2\pi)^{3}\delta^{(3)}(\vec{p}-\vec{q}). \label{com2}
\end{align}
The commutators vanish for every other pair.

The momenta given in \cref{momentum1,momentum2} can be derived from the expansion of $\hat{\phi}$. The Hamiltonian of the system is then obtained by integrating the Hamiltonian density \eqref{HamiltonianHD1} over space and making use of the above commutators.  It reads
\begin{align}
H=\int d^{3}x\mathcal{H}=&\int \frac{d^{3}p}{(2\pi)^{3}}\biggl\{\sqrt{1-4m^{2}\epsilon^{2}}\;\omega_{c\vec{p}}\,\hat{c}^{\dagger}_{\vec{p}}\hat{c}_{\vec{p}}-\sqrt{1-4m^{2}\epsilon^{2}}\;\omega_{d\vec{p}}\,\hat{d}^{\dagger}_{\vec{p}}\hat{d}_{\vec{p}}\biggr\}\nonumber\\
&+\int \frac{d^{3}p}{(2\pi)^{3}}\frac{1}{2}\sqrt{1-4m^{2}\epsilon^{2}}\delta^{(3)}(\vec{0})\left[\omega_{c\vec{p}}-\omega_{d\vec{p}}\right]. \label{Hcd2}
\end{align}
We can thus notice that every ``convergent-type particle" of momentum $\vec{p}$ contributes with the energy $\sqrt{1-4m^{2}\epsilon^{2}}\;\omega_{c\vec{p}}$ to the system, whereas every ``divergent-type particle" contributes with the energy $-\sqrt{1-4m^{2}\epsilon^{2}}\;\omega_{d\vec{p}}$. Moreover, there are two divergent vacuum energy contributions of opposite signs which, in absence of gravity, can be removed by normal ordering of $H$.

Negative energy states also lead to an energy unbounded from below. Indeed, one can construct the Fock space by applying creation operators on the vacuum state $\ket{0}$ defined through: 
\begin{equation}
\hat{c}_{\vec{p}}\ket{0}=0\;\;,\;\;\hat{d}_{\vec{p}}\ket{0}=0,\;\;\;\;\;\;\forall \vec{p}\,\,.
\end{equation}
According to \cref{Hcd2}, the energy of a state with $n_{k}$ convergent-type particles of momentum $\vec{p}_{k}$ and $m_{k}$ divergent-type particles of momentum $\vec{q}_{k}$, for $k=1,\dots,N$, is then given by
\begin{equation}
E=\sqrt{1-4m^{2}\epsilon^{2}}\sum_{k=1}^{N}\left[n_{k}\omega_{c\vec{p}_{k}}-m_{k}\omega_{d\vec{q}_{k}}\right],
\end{equation}
where the vacuum energy has been eliminated through normal ordering. Even if $\ket{0}$ has energy $E=0$, we see that creating divergent-type particles out of this state lowers the energy to negative values. The energy is thus unbounded from below and there is no ground state.

\section{Notation and conventions}\label{AppendixSuperfields}

For both notation and conventions we follow \cite{Wess:1992cp}. We thus use the mostly-plus metric $(-,+,+,+)$ to raise, lower or contract Lorentz indices. To raise, lower or contract spinorial indices we use the antisymmetric tensors
\begin{equation}
\epsilon^{\alpha\beta}=\epsilon^{\dot{\alpha}\dot{\beta}}=\begin{pmatrix}
0 & -1 \\
+1 & 0 
\end{pmatrix}=-\epsilon_{\alpha\beta}=-\epsilon_{\dot{\alpha}\dot{\beta}}.
\end{equation}
Introducing the generalized Pauli matrices
\begin{equation}
(\sigma^{\mu})_{\alpha\dot{\alpha}}=\left(\mathds{1},\vec{\sigma}\right), \quad (\bar{\sigma}^{\mu})^{\dot{\alpha}\alpha}=\left(-\mathds{1},\vec{\sigma}\right)=\epsilon^{\alpha\beta}\epsilon^{\dot{\alpha}\dot{\beta}}(\sigma^{\mu})_{\beta\dot{\beta}},
\end{equation}
we show the following relation on Grassmann variables, often used in superspace computations 
\begin{equation}
\theta^{\alpha}\theta^{\beta}=-\frac{1}{2}\epsilon^{\alpha\beta}\theta\theta\;\;,\;\;\;\bar{\theta}^{\dot{\alpha}}\bar{\theta}^{\dot{\beta}}=\frac{1}{2}\epsilon^{\dot{\alpha}\dot{\beta}}\bar{\theta}\bar{\theta},
\end{equation}
\begin{equation}
\theta\sigma^{\mu}\bar{\theta}\theta\sigma^{b}\bar{\theta}=-\frac{1}{2}\theta\theta\bar{\theta}\bar{\theta}\eta^{ab},
\end{equation}
\begin{equation}
Tr\lbrace\sigma^{\mu}\bar{\sigma}^{\nu}\rbrace\equiv(\sigma^{\mu})_{\alpha\dot{\alpha}}(\bar{\sigma}^{\nu})^{\dot{\alpha}\alpha}=-2\eta^{\mu\nu}\,\,.
\end{equation}
We also write down the superspace derivatives
\begin{equation}
D_{A}\equiv(\partial_{\mu},D_{\alpha},\bar{D}^{\dot{\alpha}})\,\,,
\end{equation}
in terms of the space-time and SUSY covariant derivatives
\begin{align}
D_{\alpha}=&+\frac{\partial}{\partial\theta^{\alpha}}+i(\sigma^{\mu}\bar{\theta})_{\alpha}\frac{\partial}{\partial x^{\mu}},\nonumber\\
\bar{D}_{\dot{\alpha}}=&-\frac{\partial}{\partial\bar{\theta}^{\dot{\alpha}}}-i(\theta\sigma^{\mu})_{\dot{\alpha}}\frac{\partial}{\partial x^{\mu}}.
\end{align}
The latter satisfy the graded Lie algebra
\begin{align}
&\lbrace D_{\alpha},\bar{D}_{\dot{\alpha}}\rbrace=-2i(\sigma^{\mu})_{\alpha\dot{\alpha}}\partial_{\mu},\label{e3.9} \\
&\lbrace D_{\alpha},D_{\beta}\rbrace=\lbrace \bar{D}_{\dot{\alpha}},\bar{D}_{\dot{\beta}}\rbrace=\left[D_{\alpha},\partial_{\mu}\right]=\left[\bar{D}_{\dot{\alpha}},\partial_{\mu}\right]=0.
\end{align}
The anticommutation relation \eqref{e3.9} can be used to express space-time derivatives of a generic superfield $S$ through SUSY derivatives only:
\begin{equation}
\partial^{\mu} S = \frac{i}{4} (\bar{\sigma}^{\mu})^{\dot\alpha \beta} \{D_{\beta}, \bar{D}_{\dot{\alpha}} \}S. \label{stDvsSUSYD}
\end{equation}
We use the standard notation $D^{2}\equiv D^{\alpha}D_{\alpha}$ and the superspace identity for a superfield $S$:
\begin{equation}
\int d^4\theta S = - \frac{1}{4}\int d^2\theta \bar{D}^2 S =\frac{1}{16} D^2\bar{D}^2 S|_0 \, . \label{ibpsupefield}
\end{equation} 
In the present paper, we mostly use chiral and anti-chiral superfields, satisfying the constraints
\begin{equation}
\bar{D}_{\dot{\alpha}}\Phi=0\;\;,\;\;\;D_{\alpha}\bar{\Phi}=0. \label{chiral}
\end{equation}
\Cref{e3.9,chiral} shows that a chiral superfield $\Phi$ satisfies:
\begin{align}
&\bar{D}^2D^2\Phi=-16\Box\Phi, \label{boxPhi} \\
&D^{\alpha}\bar{D}_{\dot{\alpha}}D_{\alpha}\Phi=-\frac{1}{2}\bar{D}_{\dot{\alpha}}{D}^2\Phi\label{superderivformula}.
\end{align}
The bosonic chiral and anti-chiral component expansions of chiral superfields read:
\begin{align}
\Phi&=\phi+i\theta\sigma^{\mu}\bar{\theta}\partial_{\mu}\phi+\frac{1}{4}\theta\theta\bar{\theta}\bar{\theta}\Box \phi+\theta\theta F,\\
\bar{\Phi}&=\bar \phi-i\theta\sigma^{\mu}\bar{\theta}\partial_{\mu}\bar \phi+\frac{1}{4}\theta\theta\bar{\theta}\bar{\theta}\Box \bar \phi+\bar{\theta}\bar{\theta} \bar F ,
\end{align}
with both $\phi(x)$ and $F(x)$ complex fields.

\section{Dimension-five HD SUSY operators in components} \label{HDMSSMcomponents}

In this appendix we show explicitly that the two Lagrangians (wrong and correct) shown in \cref{MSSMcorrectionsuperpot} differ exactly by the higher order of the field redefinition used to obtain the correct one from the initial HD Lagrangian. 

The bosonic components of the  correct effective Lagrangian \eqref{LagrangianGD2} read:
\begin{align}
    \mathcal{L}
    &=\phi\Box\bar \phi + F\bar F - 4\epsilon ^2 \sigma^2 \left( W'\Box\bar W' + F W'' \bar F \bar W''\right) \nonumber\\
    &\hspace{0.5cm}+\left\{FW'-2\epsilon\sigma F W''W'+2\epsilon^2\sigma^2 F( W'^2 W'')'\right\}+ h.c. + O(\epsilon^3).
\end{align}
This Lagrangian is linear in the auxiliary field $F$. Its equation of motion is solved by:
\begin{equation}
    \bar{F}= - W' + 2 \epsilon \sigma W''W'  - 2\epsilon^2\sigma^2 ({W'}^2W'')' +4\sigma^2 \epsilon^2 W'|W''|^2+ O(\epsilon^3). \label{Fsol}
\end{equation}
Putting back the solution for the auxiliary field in the total component Lagrangian thus gives:
\begin{align}
\mathcal{L}&=\phi\Box\bar \phi - 4\epsilon ^2 \sigma^2  W'\Box\bar W' - F \bar F (1-4\epsilon^2\sigma^2 |W''|^2) \nonumber\\
&=\phi\Box\bar \phi - 4\epsilon ^2 \sigma^2  W'\Box\bar W' -W'\bar W' + 2 \epsilon \sigma \big( W''|W'|^2+\bar W'' |W'|^2\big)   - 8 \epsilon^2\sigma^2 |W'|^2|W''|^2 \nonumber\\ 
&\hspace{1cm}  -2\epsilon^2\sigma^2 \bar W' \big(W'^2 W''\big)' -2\epsilon^2\sigma^2  W' \big(\bar W'^2 \bar W''\big)' + O(\epsilon^3). \label{Lcase2compoGD}
\end{align}
The same Lagrangian is obtained starting from the initial HD Lagrangian \eqref{Lcase2} in components, solving for the auxiliary field $F$ and performing the field redefinition:
\begin{equation}
\phi \rightarrow\phi + \delta \phi \equiv \phi + 2\epsilon\sigma W' + 2 \epsilon^2\sigma^2 \Box \phi + 2\epsilon^2\sigma^2 ( 2W''W'+\bar W'' W') + O(\epsilon^3). \label{fieldredefJLM}
\end{equation}
This field redefinition is actually identical to the superspace one made by the authors of \cite{Dudas:2015vka} to derive the Lagrangian \eqref{LagrangianGD2}.

Note that in the above Lagrangians, we did more than just removing the higher-derivative terms. Indeed, kinetic terms such as $W'\Box \bar \phi$, which are not higher-derivative terms, have also be replaced by on-shell equivalent terms. 

We now look at the expansion of the wrong Lagrangian \eqref{Lwrong} in components. After solving for the auxiliary field, the scalar field Lagrangian reads: 
\begin{align}
    \mathcal{L}_{wrong}
    &=\phi\Box\bar \phi - W'\bar W'  - 2\epsilon \sigma \big(|W'|^2  W'' +|W'|^2 \bar W''\big) -8\epsilon^2\sigma^2 W'\Box\bar W' \nonumber\\
    &\hspace{1cm} -4\epsilon^2\sigma^2 \big(|W'|^2 {W''}^2+|W'|^2 \bar{W''}^2\big) - 4\epsilon^2\sigma^2 |W''|^2 |W'|^2 + O(\epsilon^3). \label{LfinalJLM}
\end{align} 
This result is inequivalent to \cref{Lcase2compoGD}. According to the discussion at the end of \cref{fieldredefs}, we expect the difference between the two Lagrangians to come from terms appearing at second order in $\delta \phi$ (of the field redefinition \eqref{fieldredefJLM}), that are not of order $O(\epsilon^3)$ but rather $O(\epsilon^2)$. The difference should thus come from  terms of order $O(\epsilon^2)$ generated by expansion at second order of the initial Lagrangian under the field redefinition:
\begin{equation}
\phi \rightarrow \phi + \delta\phi=\phi + 2\epsilon\sigma W'.
\end{equation}
 They are:
\begin{align}
\mathcal{L}_{\rm diff}&=4\epsilon^2\sigma^2\left(\frac{\delta^2 \mathcal{L}_0}{\delta \phi^2} {W'}^2+\frac{\delta^2 \mathcal{L}_0}{\delta \bar \phi^2}  \bar W'^2+\frac{\delta^2 \mathcal{L}_0}{\delta \phi\delta \bar \phi}   |W'|^2\right)\nonumber\\
&=4\epsilon^2\sigma^2(W'\Box\bar W' - \frac12 (W''' \bar W') W'^2 - \frac12 (\bar W''' W') W'^2 - |W'|^2|W''|^2 ). 
\end{align}
We see that the wrong Lagrangian \eqref{LfinalJLM} supplemented with $\mathcal{L}_{\rm diff}$ gives indeed exactly the correct Lagrangian \eqref{Lcase2compoGD}, obtained through component or superfield field redefinitions.

\bibliography{HDSUSY}

\providecommand{\href}[2]{#2}\begingroup\raggedright\begin{thebibliography}{10}

\bibitem{Ostrogradsky:1850fid}
M.~Ostrogradsky, \emph{{M\'emoires sur les \'equations diff\'erentielles,
  relatives au probl\`eme des isop\'erim\`etres}}, {\emph{Mem. Acad. St.
  Petersbourg} {\bf 6} (1850) 385--517}.

\bibitem{Horndeski:1974wa}
G.~W. Horndeski, \emph{{Second-order scalar-tensor field equations in a
  four-dimensional space}},
  \href{http://dx.doi.org/10.1007/BF01807638}{\emph{Int. J. Theor. Phys.} {\bf
  10} (1974) 363--384}.

\bibitem{Langlois:2015cwa}
D.~Langlois and K.~Noui, \emph{{Degenerate Higher Derivative Theories Beyond
  Horndeski: Evading the Ostrogradski Instability}},
  \href{http://dx.doi.org/10.1088/1475-7516/2016/02/034}{\emph{JCAP} {\bf 02}
  (2016) 034}, [\href{https://arxiv.org/abs/1510.06930}{{\tt 1510.06930}}].

\bibitem{Buchbinder:1994iw}
I.~L. Buchbinder, S.~Kuzenko and Z.~Yarevskaya, \emph{{Supersymmetric effective
  potential: Superfield approach}},
  \href{http://dx.doi.org/10.1016/0550-3213(94)90466-9}{\emph{Nucl. Phys. B}
  {\bf 411} (1994) 665--692}.

\bibitem{Pickering:1996he}
A.~Pickering and P.~C. West, \emph{{The One loop effective superpotential and
  nonholomorphicity}},
  \href{http://dx.doi.org/10.1016/0370-2693(96)00702-2}{\emph{Phys. Lett. B}
  {\bf 383} (1996) 54--62}, [\href{https://arxiv.org/abs/hep-th/9604147}{{\tt
  hep-th/9604147}}].

\bibitem{Burgess:2014lwa}
C.~P. Burgess and M.~Williams, \emph{{Who You Gonna Call? Runaway Ghosts,
  Higher Derivatives and Time-Dependence in EFTs}},
  \href{http://dx.doi.org/10.1007/JHEP08(2014)074}{\emph{JHEP} {\bf 08} (2014)
  074}, [\href{https://arxiv.org/abs/1404.2236}{{\tt 1404.2236}}].

\bibitem{Khoury:2010gb}
J.~Khoury, J.-L. Lehners and B.~Ovrut, \emph{{Supersymmetric P(X,$\phi$) and
  the Ghost Condensate}},
  \href{http://dx.doi.org/10.1103/PhysRevD.83.125031}{\emph{Phys. Rev. D} {\bf
  83} (2011) 125031}, [\href{https://arxiv.org/abs/1012.3748}{{\tt
  1012.3748}}].

\bibitem{Koehn:2012ar}
M.~Koehn, J.-L. Lehners and B.~A. Ovrut, \emph{{Higher-Derivative Chiral
  Superfield Actions Coupled to ${\mathcal{N}}\!=1$ Supergravity}},
  \href{http://dx.doi.org/10.1103/PhysRevD.86.085019}{\emph{Phys. Rev. D} {\bf
  86} (2012) 085019}, [\href{https://arxiv.org/abs/1207.3798}{{\tt
  1207.3798}}].

\bibitem{Koehn:2012np}
M.~Koehn, J.-L. Lehners and B.~A. Ovrut, \emph{{Dbi Inflation in
  ${\mathcal{N}}\!=1$ Supergravity}},
  \href{http://dx.doi.org/10.1103/PhysRevD.86.123510}{\emph{Phys. Rev. D} {\bf
  86} (2012) 123510}, [\href{https://arxiv.org/abs/1208.0752}{{\tt
  1208.0752}}].

\bibitem{Farakos:2012qu}
F.~Farakos and A.~Kehagias, \emph{{Emerging Potentials in Higher-Derivative
  Gauged Chiral Models Coupled to N=1 Supergravity}},
  \href{http://dx.doi.org/10.1007/JHEP11(2012)077}{\emph{JHEP} {\bf 11} (2012)
  077}, [\href{https://arxiv.org/abs/1207.4767}{{\tt 1207.4767}}].

\bibitem{Ciupke:2015msa}
D.~Ciupke, J.~Louis and A.~Westphal, \emph{{Higher-Derivative Supergravity and
  Moduli Stabilization}},
  \href{http://dx.doi.org/10.1007/JHEP10(2015)094}{\emph{JHEP} {\bf 10} (2015)
  094}, [\href{https://arxiv.org/abs/1505.03092}{{\tt 1505.03092}}].

\bibitem{Antoniadis:2007xc}
I.~Antoniadis, E.~Dudas and D.~M. Ghilencea, \emph{{Supersymmetric Models with
  Higher Dimensional Operators}},
  \href{http://dx.doi.org/10.1088/1126-6708/2008/03/045}{\emph{JHEP} {\bf 03}
  (2008) 045}, [\href{https://arxiv.org/abs/0708.0383}{{\tt 0708.0383}}].

\bibitem{Antoniadis:2008es}
I.~Antoniadis, E.~Dudas, D.~M. Ghilencea and P.~Tziveloglou, \emph{{MSSM with
  Dimension-five Operators (MSSM(5))}},
  \href{http://dx.doi.org/10.1016/j.nuclphysb.2008.09.019}{\emph{Nucl. Phys. B}
  {\bf 808} (2009) 155--184}, [\href{https://arxiv.org/abs/0806.3778}{{\tt
  0806.3778}}].

\bibitem{Antoniadis:2009rn}
I.~Antoniadis, E.~Dudas, D.~M. Ghilencea and P.~Tziveloglou, \emph{{MSSM Higgs
  with dimension-six operators}},
  \href{http://dx.doi.org/10.1016/j.nuclphysb.2010.01.010}{\emph{Nucl. Phys. B}
  {\bf 831} (2010) 133--161}, [\href{https://arxiv.org/abs/0910.1100}{{\tt
  0910.1100}}].

\bibitem{Jaen:1986iz}
X.~Jaen, J.~Llosa and A.~Molina, \emph{{A Reduction of Order Two for Infinite
  Order Lagrangians}},
  \href{http://dx.doi.org/10.1103/PhysRevD.34.2302}{\emph{Phys. Rev. D} {\bf
  34} (1986) 2302}.

\bibitem{Antoniadis:2019xwa}
I.~Antoniadis, H.~Jiang and O.~Lacombe, \emph{{Note on supersymmetric
  Dirac-Born-Infeld action with Fayet-Iliopoulos term}},
  \href{http://dx.doi.org/10.1007/JHEP05(2020)111}{\emph{JHEP} {\bf 05} (2020)
  111}, [\href{https://arxiv.org/abs/1912.12627}{{\tt 1912.12627}}].

\bibitem{Solomon:2017nlh}
A.~R. Solomon and M.~Trodden, \emph{{Higher-derivative operators and effective
  field theory for general scalar-tensor theories}},
  \href{http://dx.doi.org/10.1088/1475-7516/2018/02/031}{\emph{JCAP} {\bf 02}
  (2018) 031}, [\href{https://arxiv.org/abs/1709.09695}{{\tt 1709.09695}}].

\bibitem{deUrries:1998obu}
F.~J. de~Urries and J.~Julve, \emph{{Ostrogradski Formalism for Higher
  Derivative Scalar Field Theories}},
  \href{http://dx.doi.org/10.1088/0305-4470/31/33/006}{\emph{J. Phys. A} {\bf
  31} (1998) 6949--6964}, [\href{https://arxiv.org/abs/hep-th/9802115}{{\tt
  hep-th/9802115}}].

\bibitem{Georgi:1991ch}
H.~Georgi, \emph{{On-Shell Effective Field Theory}},
  \href{http://dx.doi.org/10.1016/0550-3213(91)90244-R}{\emph{Nucl. Phys. B}
  {\bf 361} (1991) 339--350}.

\bibitem{Weinberg:2008hq}
S.~Weinberg, \emph{{Effective Field Theory for Inflation}},
  \href{http://dx.doi.org/10.1103/PhysRevD.77.123541}{\emph{Phys. Rev. D} {\bf
  77} (2008) 123541}, [\href{https://arxiv.org/abs/0804.4291}{{\tt
  0804.4291}}].

\bibitem{Burgess:2007pt}
C.~P. Burgess, \emph{{Introduction to Effective Field Theory}},
  \href{http://dx.doi.org/10.1146/annurev.nucl.56.080805.140508}{\emph{Ann.
  Rev. Nucl. Part. Sci.} {\bf 57} (2007) 329--362},
  [\href{https://arxiv.org/abs/hep-th/0701053}{{\tt hep-th/0701053}}].

\bibitem{Grosse-Knetter:1993tae}
C.~Grosse-Knetter, \emph{{Effective Lagrangians with Higher Derivatives and
  Equations of Motion}},
  \href{http://dx.doi.org/10.1103/PhysRevD.49.6709}{\emph{Phys. Rev. D} {\bf
  49} (1994) 6709--6719}, [\href{https://arxiv.org/abs/hep-ph/9306321}{{\tt
  hep-ph/9306321}}].

\bibitem{Gong:2014rna}
J.-O. Gong, M.-S. Seo and S.~Sypsas, \emph{{Higher Derivatives and Power
  Spectrum in Effective Single Field Inflation}},
  \href{http://dx.doi.org/10.1088/1475-7516/2015/03/009}{\emph{JCAP} {\bf 03}
  (2015) 009}, [\href{https://arxiv.org/abs/1407.8268}{{\tt 1407.8268}}].

\bibitem{Wess:1992cp}
J.~Wess and J.~Bagger, \emph{{Supersymmetry and Supergravity}}.
\newblock Princeton University Press, Princeton, NJ, USA, 1992.

\bibitem{Ciupke:2016agp}
D.~Ciupke, \emph{{Scalar Potential from Higher Derivative $\mathcal{N} = 1$
  Superspace}},  \href{https://arxiv.org/abs/1605.00651}{{\tt 1605.00651}}.

\bibitem{Bielleman:2016grv}
S.~Bielleman, L.~E. Ibanez, F.~G. Pedro, I.~Valenzuela and C.~Wieck, \emph{{The
  Dbi Action, Higher-Derivative Supergravity, and Flattening Inflaton
  Potentials}}, \href{http://dx.doi.org/10.1007/JHEP05(2016)095}{\emph{JHEP}
  {\bf 05} (2016) 095}, [\href{https://arxiv.org/abs/1602.00699}{{\tt
  1602.00699}}].

\bibitem{Bielleman:2016olv}
S.~Bielleman, L.~E. Ibanez, F.~G. Pedro, I.~Valenzuela and C.~Wieck,
  \emph{{Higgs-Otic Inflation and Moduli Stabilization}},
  \href{http://dx.doi.org/10.1007/JHEP02(2017)073}{\emph{JHEP} {\bf 02} (2017)
  073}, [\href{https://arxiv.org/abs/1611.07084}{{\tt 1611.07084}}].

\bibitem{Dudas:2015vka}
E.~Dudas and D.~M. Ghilencea, \emph{{Effective Operators in Susy, Superfield
  Constraints and Searches for a UV Completion}},
  \href{http://dx.doi.org/10.1007/JHEP06(2015)124}{\emph{JHEP} {\bf 06} (2015)
  124}, [\href{https://arxiv.org/abs/1503.08319}{{\tt 1503.08319}}].

\bibitem{Delgado:2023ivp}
A.~Delgado, A.~Martin and R.~Wang, \emph{{Counting operators in N = 1
  supersymmetric gauge theories}},
  \href{http://dx.doi.org/10.1007/JHEP07(2023)081}{\emph{JHEP} {\bf 07} (2023)
  081}, [\href{https://arxiv.org/abs/2305.01736}{{\tt 2305.01736}}].

\bibitem{Antoniadis:1997eg}
I.~Antoniadis, S.~Ferrara, R.~Minasian and K.~S. Narain, \emph{{R**4 couplings
  in M and type II theories on Calabi-Yau spaces}},
  \href{http://dx.doi.org/10.1016/S0550-3213(97)00572-5}{\emph{Nucl. Phys. B}
  {\bf 507} (1997) 571--588}, [\href{https://arxiv.org/abs/hep-th/9707013}{{\tt
  hep-th/9707013}}].

\bibitem{Becker:2002nn}
K.~Becker, M.~Becker, M.~Haack and J.~Louis, \emph{{Supersymmetry breaking and
  alpha-prime corrections to flux induced potentials}},
  \href{http://dx.doi.org/10.1088/1126-6708/2002/06/060}{\emph{JHEP} {\bf 06}
  (2002) 060}, [\href{https://arxiv.org/abs/hep-th/0204254}{{\tt
  hep-th/0204254}}].

\bibitem{Grimm:2017okk}
T.~W. Grimm, K.~Mayer and M.~Weissenbacher, \emph{{Higher derivatives in Type
  II and M-theory on Calabi-Yau threefolds}},
  \href{http://dx.doi.org/10.1007/JHEP02(2018)127}{\emph{JHEP} {\bf 02} (2018)
  127}, [\href{https://arxiv.org/abs/1702.08404}{{\tt 1702.08404}}].

\bibitem{Cicoli:2016chb}
M.~Cicoli, D.~Ciupke, S.~de~Alwis and F.~Muia, \emph{{$\alpha'$ Inflation:
  moduli stabilisation and observable tensors from higher derivatives}},
  \href{http://dx.doi.org/10.1007/JHEP09(2016)026}{\emph{JHEP} {\bf 09} (2016)
  026}, [\href{https://arxiv.org/abs/1607.01395}{{\tt 1607.01395}}].

\bibitem{Cicoli:2023njy}
M.~Cicoli, M.~Licheri, P.~Piantadosi, F.~Quevedo and P.~Shukla, \emph{{Higher
  derivative corrections to string inflation}},
  \href{http://dx.doi.org/10.1007/JHEP02(2024)115}{\emph{JHEP} {\bf 02} (2024)
  115}, [\href{https://arxiv.org/abs/2309.11697}{{\tt 2309.11697}}].

\bibitem{Weldon:2003by}
H.~A. Weldon, \emph{{Quantization of higher-derivative field theories}},
  \href{http://dx.doi.org/10.1016/S0003-4916(03)00070-8}{\emph{Annals Phys.}
  {\bf 305} (2003) 137--150}.

\bibitem{Woodard:2015zca}
R.~P. Woodard, \emph{{Ostrogradsky's Theorem on Hamiltonian Instability}},
  \href{http://dx.doi.org/10.4249/scholarpedia.32243}{\emph{Scholarpedia} {\bf
  10} (2015) 32243}, [\href{https://arxiv.org/abs/1506.02210}{{\tt
  1506.02210}}].

\end{thebibliography}\endgroup
\bibliographystyle{JHEP}
  
\end{document}